\documentclass[superscriptaddress,nofootinbib,notitlepage,10pt]{revtex4-1}

\usepackage{amsfonts}
\usepackage[bookmarksopen,colorlinks]{hyperref}
\usepackage{graphicx}
\usepackage{dcolumn}
\usepackage{bm}
\usepackage{amssymb}
\usepackage{amsmath,mleftright}
\usepackage{xparse}
\usepackage{comment}
\usepackage{enumitem}
\usepackage[dvipsnames]{xcolor}

\begin{document}

\title{Pseudoinvariance and the extra degree of freedom in $f(T)$ gravity}
\author{Rafael Ferraro}
\email{ferraro@iafe.uba.ar} \affiliation{Instituto de Astronom\'ia y
F\'isica del Espacio (IAFE, CONICET-UBA), Casilla de Correo 67,
Sucursal 28, 1428 Buenos Aires, Argentina.}
\affiliation{Departamento de F\'isica, Facultad de Ciencias Exactas
y Naturales, Universidad de Buenos Aires, Ciudad Universitaria,
Pabell\'on I, 1428 Buenos Aires, Argentina.}

\author{Mar\'ia Jos\'e Guzm\'an}
\email{maria.j.guzman.m@gmail.com} \affiliation{Departamento de
F\'isica y Astronom\'ia, Facultad de Ciencias, Universidad de La
Serena, Av. Juan Cisternas 1200, 1720236 La Serena, Chile.}

\begin{abstract}
Nonlinear generalizations of teleparallel gravity entail the modification of a Lagrangian that is pseudoinvariant under local Lorentz transformations of the tetrad field. This procedure consequently leads to the loss of the local pseudoinvariance and the appearance of additional degrees of freedom (d.o.f.). The constraint structure of $f(T)$ gravity suggests the existence of one extra d.o.f. when compared with general relativity, which should describe some aspect of the orientation of the tetrad. The purpose of this article is to better understand the nature of this extra d.o.f. by means of a toy model that mimics essential features of $f(T)$ gravity. We find that the nonlinear modification of a Lagrangian $L$ possessing a local rotational pseudoinvariance produces two types of solutions. In one case the original gauge-invariant variables --the analogue of the metric in teleparallelism-- evolve like when governed by the (nondeformed) Lagrangian $L$; these solutions are characterized by a (selectable) constant value of its Lagrangian, which is the manifestation of the extra d.o.f. In the other case, the solutions do contain new dynamics for the original gauge-invariant variables, but the extra d.o.f. does not materialize because the Lagrangian remains invariant on-shell. Coming back to $f(T)$ gravity, the first case includes solutions where the torsion scalar $T$ is a constant, to be chosen at the initial conditions (extra d.o.f.), and no new dynamics for the metric is expected. The latter case covers those solutions displaying a genuine modified gravity; $T$ is not a constant, but it is (on-shell) invariant under Lorentz transformations depending only on time. Both kinds of $f(T)$ solutions are exemplified in a flat Friedmann-Lemaître-Robertson-Walker universe. Finally, we present a toy model for a higher-order Lagrangian with rotational invariance [analogous to $f(R)$ gravity] and derive its constraint structure and number of d.o.f. 
\end{abstract}

\maketitle
\tableofcontents

\section{Introduction}

\label{Intro} The common notion that gravity can only be represented through
the curvature of spacetime has being challenged by at least two different
approaches, where either the torsion or the nonmetricity provide physically
and mathematically equivalent versions of general relativity (GR). These two
theories correspond to the teleparallel equivalent of general relativity
(TEGR) \cite{Ald13} and the symmetric teleparallel equivalent of general
relativity (STEGR)\cite{Adak:2004uh,Adak:2005cd}, and their dynamical
variables are the torsion tensor and the nonmetricity tensor, respectively.
The description of general relativity in terms of curvature, torsion and
non-metricity has incidentally being called the \textquotedblleft
geometrical trinity of gravity\textquotedblright\ \cite%
{BeltranJimenez:2017tkd,BeltranJimenez:2019tjy}, and it consists in an
intriguing starting point to formulate extensions of Einstein's gravity. The
TEGR as a starting point for building extensions to general relativity has
gained wide attention in the recent years, particularly for its versatility
to predict novel consequences in the realm of cosmology, giving rise to the $%
f(T)$ gravity paradigm \cite{Ferraro:2006jd,Bengochea:2008gz}, where $T$ is
the torsion scalar. Equivalently, in STEGR the nonmetricity
scalar $Q$ is used, giving rise to the very recent $f(Q)$ theories of gravity \cite%
{Jimenez:2019ovq}.

Recent interest has emerged for understanding the issue of the number and
nature of the degrees of freedom in modified gravity theories based on a
teleparallel framework. Some early attempts to understand $f(T)$ gravity as
TEGR plus a minimally coupled scalar field through conformal transformations
were documented in Refs. \cite{Yang:2010ji,Wright:2016ayu}, where it was shown that
it is not possible to cleanly obtain a teleparallel Einstein frame, due to
the appearance of Lorentz-breaking terms. However, later it was shown
through a full Hamiltonian analysis, that $f(T)$ gravity has a unique extra
degree of freedom (d.o.f.) \cite{Ferraro:2018tpu}, which consequently cannot
be attributed to a conformal field redefinition of the theory. In this
regard, recently disformal transformations were studied in order to
obtain a clean isolation of such an extra d.o.f., but these efforts have been
unsuccessful \cite{Golovnev:2019kcf}. Other attempts to understand the issue 
of the d.o.f. worth considering in this discussion are the studies
of the linearized approximation around Minkowski spacetime \cite%
{Bamba:2013ooa,Cai:2018rzd,Hohmann:2018jso}, which do not show the extra
d.o.f. Also, propagating modes do not appear in linear cosmological
perturbations around a spatially flat Friedmann-Lemaître-Roberston-Walker universe \cite%
{Chen:2010va,Li:2011wu,Izumi:2012qj,Golovnev:2018wbh,Sahlu:2019bug,Toporensky:2019mlg,Golovnev:2020aon}. In
the light of the perturbative analysis, there are concerns in the community
regarding the pathological behavior and strong coupling problem in $f(T)$
gravity \cite%
{Izumi:2012qj,Golovnev:2018wbh,Koivisto:2018loq,Jimenez:2019tkx,Jimenez:2019ghw}%
. The disappearance of degrees of freedom at the perturbative level is a
behavior shared with other modified gravitational theories such as massive,
bimetric and Ho\v{r}ava gravities. Nonetheless, an important distinction
between these theories and $f(T)$ gravity is that the former use the metric
as the dynamical field; in contrast $f(T)$ gravity is a tetrad-based
physical theory.  The extra d.o.f. can be roughly interpreted
as a scalar field that has a role in selecting preferred reference frames
that are solutions of the equations of motion \cite{Ferraro:2018axk},
exhibiting in this way the loss of local Lorentz invariance (LLI). So, it is still unclear if it should dynamically manifest at the perturbative level, putting in doubt concerns about the strong coupling problem. 

Another road to understanding the important matter of the lack of LLI in these
theories comes from the analysis of pseudoinvariance in TEGR. It is widely
known that TEGR is a pseudoinvariant (also called \textit{quasi}-invariant
\cite{Obukhov:2006fy}) theory under local Lorentz transformations (LLT) on
the tetrad field. This means that the TEGR Lagrangian changes by a boundary
term under LLT, or in other words, the difference between the Ricci scalar $R
$ from GR and the torsion scalar $T$ in TEGR is a boundary term. Therefore,
in the nonlinear modification of the TEGR Lagrangian, we cannot integrate
out this boundary term, giving rise to the modification of a
pseudoinvariant system. Boundary terms are very common in GR, such as
topological invariants that are nontrivial in higher dimensions or the
Gibbons-Hawking-York term,\footnote{%
It has been claimed that the surface term from TEGR has the same
contribution, once varied, as the Gibbons-Hawking-York term,
erasing in the same way the unwanted contributions to the Einstein
equations of motion when spacetime boundaries are considered
\cite{Oshita:2017nhn}} but the nonlinear modifications of these
terms are not commonly used for model building. In this regard, we
have a very unique case of modified pseudoinvariance in
modifications to gravity based on the teleparallel formalism. Our
aim is to analyze the properties of pseudoinvariant systems and their 
nonlinear modifications through toy models, which will be very
helpful to understand the disappearance of the extra d.o.f. in
$f(T)$ gravity for some solutions and its general behavior.

This work is organized as follows. In Sec. \ref{sec:mtg} we introduce the
basic concepts and definitions of teleparallel and modified teleparallel
gravity. In Sec. \ref{sec:toymodel} we present the Hamiltonian analysis
of a toy model with rotational pseudoinvariance, and the analysis of the
nonlinear modification of it. We compare the outcome and generic features
of the toy model with the $f(T)$ gravity case in Sec. \ref{sec:ftdof},
and classify a couple of qualitatively different cosmological backgrounds.
In Sec. \ref{sec:rotinv} we display a different toy model that shares
some features with $f(R)$ gravity. Section \ref{sec:concl} is devoted to the
conclusions.

\section{Teleparallel and modified teleparallel gravity}

\label{sec:mtg}

\subsection{Teleparallel geometry}

We begin by introducing the basic notation and main expressions for
understanding the teleparallel formalism. Let us consider a manifold $%
\mathcal{M}$, a basis $\{\mathbf{e}_{a}\}$ in the tangent space $T_{p}(%
\mathcal{M})$, and the dual basis $\{\mathbf{E}^{a}\}$ in the cotangent
space $T_{p}^{\ast}(\mathcal{M})$. This pair of basis/cobasis accomplishes $%
\mathbf{E}^{a}(\mathbf{e} _{b})=\delta _{b}^{a}$. When expanded in a
coordinate basis as $\mathbf{e}_{a}=e_{a}^{\mu }~\partial _{\mu }$ and $%
\mathbf{E}^{a}=E_{\mu }^{a}\ dx^{\mu }$, the duality relationship looks like
\begin{equation}
E_{\mu }^{a}\ e_{b}^{\ \mu }~=~\delta _{b}^{a}~,\ \ \ \ \ e_{a}^{\mu
}~E_{\nu }^{a}~=~\delta _{\nu }^{\mu }.  \label{duality}
\end{equation}%
Our notation is such that greek letters $\mu ,\nu ,...=0,...,n-1$ represent
spacetime coordinate indices, and latin letters $a,b,...,g,h=0,...,n-1$ are
for Lorentzian tangent space indices. A \textit{vielbein} (vierbein o tetrad
in $n=4$ dimensions) is a basis that encodes the metric structure of the
spacetime through the expression
\begin{equation}
\mathbf{g~}=~\eta _{ab}\ \mathbf{E}^{a}\otimes \mathbf{E}^{b}
\label{metric1}
\end{equation}%
[$\eta _{ab}=\text{diag} (1,-1,-1,-1)$ is the Minkowski symbol]. This allows
to write
\begin{equation}
\mathbf{E}^{a}\cdot \mathbf{E}^{b}~=~\mathbf{g}(\mathbf{\mathbf{E}}^{a},%
\mathbf{\mathbf{E}}^{b})\mathbf{~}=~\eta _{ab},
\end{equation}%
which indicates that the vielbein is an orthonormal basis. In component
notation, the former expressions are written as
\begin{equation}
g_{\mu \nu }~=~\eta _{ab}~E_{\mu }^{a}~E_{\nu }^{b},\hspace{0.25in}\eta
_{ab}~=~g_{\mu \nu }~e_{a}^{\mu }~e_{b}^{\nu },  \label{metric2}
\end{equation}%
from which the relation between the metric volume and the determinant of the
matrix $E_{\mu }^{a}$ can be derived, giving
\begin{equation}
\sqrt{|g|}~=\det [E_{\mu }^{a}]~\doteq ~E.  \label{volume}
\end{equation}

TEGR comes from the formulation of a dynamical theory of spacetime geometry for the vielbein
field, encoding the metric structure of spacetime. The Lagrangian density
for TEGR is
\begin{equation}
L~=~E\ T,  \label{TEGR}
\end{equation}%
where $T$ is the torsion scalar or Weitzenb\"{o}ck invariant,
\begin{equation}
T~\doteq ~T_{\ ~\mu \nu }^{\rho }\ S_{\rho }^{\ ~\mu \nu },  \label{Tscalar}
\end{equation}%
which is made up of the \textit{torsion tensor}
\begin{equation}
T_{\hspace{0.05in}\,\nu \rho }^{\mu }~\doteq ~e_{a}^{\ \mu }~(\partial _{\nu
}E_{\rho }^{a}-\partial _{\rho }E_{\nu }^{a}),  \label{TorsionTensor}
\end{equation}%
and the so-called \textit{superpotential}
\begin{equation}
S_{\rho }^{\ \mu \nu }~\doteq ~\dfrac{1}{2}~\left( K_{\hspace{0.05in}\hspace{%
0.05in}\rho }^{\mu \nu }+T^{\mu }~\delta _{\rho }^{\nu }-T^{\nu }~\delta
_{\rho }^{\mu }\right),
\end{equation}%
where $T^{\mu }\doteq T_{\lambda }^{\ \lambda \mu }$ is the \textit{torsion
vector}. In the latter, we define the contortion tensor as
\begin{equation}
K_{\hspace{0.05in}\hspace{0.05in}\rho }^{\mu \nu }~\doteq ~\frac{1}{2}%
~(T_{\rho }^{\hspace{0.05in}\mu \nu }-T_{\hspace{0.05in}\hspace{0.05in}\rho
}^{\mu \nu }+T_{\hspace{0.05in}\hspace{0.05in}\rho }^{\nu \mu }),
\label{contorsion}
\end{equation}%
which is the difference between the Levi-Civita connection and a general
connection. The field strength (\ref{TorsionTensor}) is the torsion
associated with the Weitzenb\"{o}ck connection $\Gamma _{\nu \rho }^{\mu
}\doteq e_{a}^{\ \mu }\,\partial _{\nu }E_{\rho }^{a}$. The Weitzenb\"{o}ck
connection is the simplest choice that cancels out the Riemann tensor,
rendering a curvatureless spacetime where the parallel transport does not
depend on the path: it is absolute. However, other choices for the
connection are possible. A modern summary and criticism of these approaches
can be found in Ref. \cite{Bejarano:2019fii}. The equations of motion for the
Lagrangian (\ref{TEGR}) are obtained by varying $L$ with respect to the
tetrad field; they are
\begin{equation}
4~e~\partial _{\mu }(E~e_{a}^{\lambda }~S_{\lambda }^{\ \mu \nu
})+4~e_{a}^{\lambda }~T_{\ \mu \lambda }^{\rho }~S_{\rho }^{\ \mu \nu
}-e_{a}^{\nu }~T=-2\kappa ~e_{a}^{\lambda }~\mathcal{T}_{\lambda }^{\ \nu },
\label{TEGReom}
\end{equation}%
where $\mathcal{T}_{\lambda }^{\ \nu }$ is the energy-momentum
tensor coming from a matter field. Equation \eqref{TEGReom} can be
proved to be equivalent to the Einstein equations when written in terms
of the metric tensor. TEGR is equivalent to GR 
not only in this sense, but also at the level of the Lagrangians.
This is because the torsion scalar $T$ and the Levi-Civita scalar
curvature $R$ are related by a boundary term
\begin{equation}
R=-T+2~e~\partial _{\mu }(E~T^{\mu })\ ,  \label{bterm}
\end{equation}%
which is integrated out once in the action, yielding the equivalence between the 
TEGR and GR Lagrangians.

\subsection{Modified teleparallel gravity}

\label{f(T)}

If our starting point to describe the gravitational interactions is the TEGR
Lagrangian, then the simplest way to a theory of modified gravity is to
replace the TEGR Lagrangian by a nonlinear function of it, in the same way that $f(R)$ gravity 
is the simplest generalization of GR. If we try to deform
gravity in this way, we can define the following action:
\begin{equation}
S=\dfrac{1}{2\kappa }\int d^{4}x~E~(f(T)+L_{m}[E_{\mu }^{a}])\ ,
\end{equation}%
where $L_{m}$ is a Lagrangian for matter. The dynamical equations of motion
of this action are found by varying in terms of the tetrad field. It is
obtained that
\begin{equation}
4~e~\partial _{\mu }(f^{\prime }(T)~E~e_{a}^{\lambda }~S_{\lambda }^{\ \mu
\nu })+4~f^{\prime }(T)~e_{a}^{\lambda }~T_{\ \mu \lambda }^{\sigma
}~S_{\sigma }^{\ \mu \nu }-e_{a}^{\nu }~f(T)=-2\kappa ~e_{a}^{\lambda }~%
\mathcal{T}_{\lambda }^{\ \nu }\ .  \label{eomft}
\end{equation}%
The equations of motion \eqref{eomft} possess an unusual feature: while they
are invariant under \textit{global} Lorentz transformations of the tetrad
field, they are sensitive to the \textit{local} orientation of the tetrad.
This means that they endow the spacetime with preferred parallelizations, which
relate each other through a subset of LLT \cite%
{Ferraro:2014owa}. The breakdown of the LLI is
irrelevant for the metric, since the components of the metric tensor are not
affected by either global or local Lorentz transformations of the tetrad
field. Then this loss of LLI is not a proper Lorentz violation in the sense
of other explicitly Lorentz-breaking gravitational theories, but implies the
existence of an extra degree of freedom \cite{Ferraro:2018tpu} that could be
only detected through interactions of matter with the tetrad field instead
of the metric.

The growing interest in $f(T)$ gravity mainly lies in its success in the
cosmological arena. In fact, a Born-Infeld-like $f(T)$ is able to smooth
spacetime singularities, leading to a maximum attainable Hubble factor in
the early Universe, and so driving an inflationary epoch without the need of
an inflaton field \cite{Ferraro:2006jd}. At the far end, the theory can
explain the accelerated expansion of the Universe by means of a power law in
the torsion scalar. In this work we are interested in understanding how the
extra degree of freedom of $f(T)$ gravity manifests itself in simple flat
FLRW cosmological backgrounds; we present a couple of solutions of this kind
in what comes next.

\subsection{Branching of cosmological solutions}

Recently it has been noticed that two different types of solutions can be
obtained when using $f(T)$ gravity in the context of flat FLRW geometries,
which present qualitatively different values for the torsion scalar. On the
one hand, the simplest and best-known solution is \cite%
{Ferraro:2006jd,Bengochea:2008gz}
\begin{equation}
\mathbf{E}^{0}=\mathbf{dt},\ \ \ \ \mathbf{E}^{1}=a(t)~\mathbf{dx},\ \ \ \
\mathbf{E}^{2}=a(t)~\mathbf{dy},\ \ \ \ \mathbf{E}^{3}=a(t)~\mathbf{dz},
\label{cosmotetrad1}
\end{equation}%
which easily proves to be a solution of the system of equations \eqref{eomft}%
. The torsion scalar for this solution is
\begin{equation}
T=-6H^{2}=-6\left( \frac{\dot{a}}{a}\right) ^{2},
\end{equation}%
and the scale factor $a(t)$ satisfies the dynamical equations \footnote{%
Incidentally, notice that these equations are invariant under the change $%
f(T)\longrightarrow f(T)+A\sqrt{T}$.} coming from replacing Eq. \eqref{cosmotetrad1} in Eq. \eqref{eomft}, giving
\begin{equation}
-2 T^{\frac{3}{2}}\dfrac{d}{dT}{\left( T^{-\frac{1}{2}} f(T)\right) }\bigg|%
_{T=-6 H^{2}}=\ 2\kappa \rho,\ \ \ \ \ \ \ -8 \dot{H} T^{\frac{1}{2}} \dfrac{d}{%
dT}{\left( T^{\frac{1}{2}}~\dfrac{df}{dT}\right) }\bigg|_{T=-6 H^{2}}=\
2\kappa (\rho +p).
\end{equation}%
The dynamics of the scale factor $a(t)$ is subject to the choice of the
function $f$; therefore this is the way the metric behavior departs from
general relativity.

On the other hand the flat FLRW geometry also allows for a family of
solutions that reads
\begin{equation}
\mathbf{E}^{0}=\cosh \lambda ~\mathbf{dt}+a(t)~\sinh \lambda ~\mathbf{dr},\
\ \ \ \mathbf{E}^{1}=\sinh \lambda ~\mathbf{dt}+a(t)~\cosh \lambda ~\mathbf{%
dr},\ \ \ \ \mathbf{E}^{2}=a(t)~r~\mathbf{d\theta },\ \ \ \ \mathbf{E}%
^{3}=a(t)~r~\sin \theta ~\mathbf{d\varphi },  \label{mccosmo}
\end{equation}%
where
\begin{equation}
\lambda (t,r)=\psi (r a(t))+\dfrac{t}{2 r a(t)}-\dfrac{r a(t)}{4}\int
(T_{o}+6H^{2})\ dt,  \label{eq:lambda}
\end{equation}%
where $T_{o}$ is a constant, and $\psi $ is an arbitrary function of the radial
distance $r~a(t)$. In this case, the torsion scalar is constant,
\begin{equation}
T=T_{o},
\end{equation}%
and the scale factor $a(t)$ satisfies the dynamical equations
\begin{equation}
6~H^{2}-T_{o}+\frac{f(T_{o})}{f^{\prime }(T_{o})}=\frac{2\kappa }{f^{\prime
}(T_{o})}~\rho,~~~~~~~~~~~~~~~~-4~\dot{H}=\frac{2\kappa }{f^{\prime
}(T_{o})}~(\rho +p),  \label{FRWeom}
\end{equation}%
which are nothing but the equations of general relativity for a cosmological
constant $\Lambda =\left( T_{o}-f(T_{o})/f^{\prime }(T_{o})\right) /2$ and
an effective Newton constant $\widehat{G}=G/f^{\prime }(T_{o})$. Then, this
other type of solution comes with an integration constant $T_{o}$ --it
appears in the radial boost governed by the function $\lambda$-- that
affects the effective values of the fundamental constants of the cosmology.
This fact was first reported in Ref.~\cite{Bejarano:2017akj} for a
vanishing value of the torsion scalar, through the null tetrad approach
developed in Ref.~\cite{Bejarano:2014bca}.

We will employ a mechanical toy model to explain why the solutions of the
original (GR) theory actually coexist with the expected new solutions of the
modified $f(T)$ theory. We are interested in knowing how many degrees of
freedom are involved in each case, and which is the remnant gauge freedom
kept by the tetrad.

\section{Modifying a mechanical system with rotational pseudoinvariance}

\label{sec:toymodel}

\subsection{Counting degrees of freedom in constrained Hamiltonian systems}

We will summarize Dirac's procedure for constrained Hamiltonian systems,\cite%
{Henneaux,Sundermeyer1,Sundermeyer2} for later use in a couple of toy
models. We consider a Lagrangian $L=L(q^{k},\dot{q}^{k})$ such that the equations
defining the canonical momenta $p_{k}=\partial L/\partial \dot{q}^{k}$
cannot be unambiguously solved for all the velocities. If so, the momenta
are not independent but there exist some relations among the $p_{k}$'s and $%
q^{k}$'s,
\begin{equation}
\phi _{\rho }(q^{k},p_{k})=0,\ \ \ \ \ \rho =1,\ldots ,P,
\label{pconstraints}
\end{equation}%
which will be called \textit{primary constraints}.\footnote{%
We will assume that the $\phi _{\rho }(q,p)$'s are independent functions.}
The constraints \eqref{pconstraints} define a subspace $\Gamma _{p}$ of the
phase space --the \textit{constraint surface}-- where the dynamics of the
system will remain confined. The \textit{primary Hamiltonian}
\begin{equation}
H_{p}\equiv H_{c}+u^{\rho }\phi _{\rho },  \label{primaryH}
\end{equation}%
is the sum of the canonical Hamiltonian $H_{c}=\dot{q}^{k}p_{k}-L(q^{k},\dot{%
q}^{k})$ and a linear combination of the primary constraints. The Lagrange
multipliers $u^{\rho }(t)$ are free functions that can be varied
independently to ensure the primary constraints. They leave $H_{p}$ with a
degree of ambiguity that comes from the fact that the velocities cannot be
uniquely solved in terms of the canonical momenta.

The condition that the primary constraints be preserved over time leads to
the following system of equations
\begin{equation}
\dot{\phi}_{\sigma }=\{\phi _{\sigma },H_{p}\}=\{\phi _{\sigma
},H_{c}\}+\{\phi _{\sigma },\phi _{\rho }\}~u^{\rho }\equiv h_{\sigma
}+C_{\sigma \rho }~u^{\rho }\overset{!}{\approx },
\end{equation}%
where $\approx 0$ means \textit{weakly} zero (i.e.\textquotedblleft zero on
the constraint surface\textquotedblright ); $h_{\sigma }$ and $C_{\sigma
\rho }$ are implicitly defined. These \textit{consistency} equations could
be accomplished by solving them for the functions $u^{\rho }$. However if $%
\det C_{\sigma \rho }\approx 0$ and $h_{\sigma }\not\approx 0$, the
consistency equations cannot be entirely solved for the functions $u^{\rho }$%
. In such a case, \textit{secondary} constraints will be needed to ensure that
the primary constraints remain weakly zero while the system evolves.%
\footnote{%
Secondary constraints will appear each time that $w_{a}^{\sigma }~h_{\sigma
}\neq 0$, where $w_{a}^{\sigma }$ is a null eigenvector of the $P\times P$
matrix $C_{\sigma \rho }$ ($w_{a}^{\sigma }~C_{\sigma \rho }\approx 0$).}
Thus, the procedure should be iterated for the consistency of the secondary
constraints, which could lead to more secondary constraints. The algorithm
finishes when the set of primary and secondary constraints,
\begin{eqnarray}
&&\phi _{\rho }\approx 0,\ \ \ \ \ \rho =1,\ldots ,P,  \notag \\
&&\phi _{\overline{\rho }}\approx 0,\ \ \ \ \ \overline{\rho }=P+1,\ldots
,P+S,
\end{eqnarray}%
can be forced to consistently evolve by merely fixing some of the Lagrange
multipliers $u^{\rho }$. We can wonder how many Lagrange multipliers will be
fixed, since some of the consistency equations could be automatically
satisfied without imposing\ any condition on the Lagrange multipliers. For
simplicity let us call $\phi _{\hat{\rho}}$, $\hat{\rho}=1,\ldots ,P+S$, the
complete set of independent constraints defining the constraint surface $%
\Gamma $. The consistency equations are
\begin{equation}
\dot{\phi}_{\hat{\rho}}=h_{\hat{\rho}}+C_{\hat{\rho}\rho }~u^{\rho }\approx
0.  \label{consistency2}
\end{equation}%
If the rank of the $S\times P$ matrix $C_{\hat{\rho}\rho }$ is $K<P$, then
there will be $P-K$ \textit{right} null eigenvectors $V_{a}^{\rho }$,
\begin{equation}
C_{\hat{\rho}\rho }~V_{a}^{\rho }~\approx ~0,~~~~~~~~~a=1,~...,~P-K.
\end{equation}%
Therefore the replacement $u^{\rho }\longrightarrow u^{\rho
}+v^{a}~V_{a}^{\rho }$, with arbitrary functions $v^{a}(t)$, will not alter
the equation \eqref{consistency2}. As a consequence, whenever the rank of $%
C_{\hat{\rho}\rho }$ is less than $P$ then it will remain an undetermined
sector in the primary Hamiltonian \eqref{primaryH} associated with the
constraints%
\begin{equation}
\phi _{a}~\equiv ~V_{a}^{\rho }~\phi _{\rho }~\approx ~0~.
\end{equation}%
Let us call \textit{first class} any phase space function $F(q,p)$ having
weakly vanishing Poisson brackets with all the constraints $\phi _{\hat{\rho}%
}$; otherwise it will be \textit{second class}. Remarkably, the constraints $%
\phi _{a}$ are first class.\footnote{$\{\phi _{\hat{\rho}},\phi
_{a}\}\approx \{\phi _{\hat{\rho}},\phi _{\rho }\}V_{a}^{\rho }=C_{\hat{\rho}%
\rho }V_{a}^{\rho }\approx 0~$. The $\phi _{a}$'s are a complete set of
first-class primary constraints, since no linearly-independent solutions to
the former equation are left on $\Gamma $.} Also $H_{p}$ is first class due
to the consistency relations. The constraints $\phi _{\hat{\rho}}$ can be
linearly combined to get a maximum number of independent first-class
constraints \textquotedblleft $\gamma _{\widetilde{A}}$.\textquotedblright 
A set of second-class constraints \textquotedblleft $\chi _{A}$%
\textquotedblright\ will complete the set of $P+S$ constraints
characterizing the constraint surface $\Gamma $. Since both $C_{a\rho }$
 and $C_{\widetilde{A}\rho }$ are weakly zero, the consistency equations
for all the first-class constraints imply nothing for the $u^{\rho }$'s. So,
let us pay attention to the consistency equations for the second-class
constraints. We notice that the square matrix $\Delta _{AB}=\{\chi _{A},\chi
_{B}\}$ must be invertible; otherwise, there would still be first-class
constraints among the $\chi _{A}$'s. Since the determinant of the
antisymmetric matrix $\Delta _{AB}$ is different from zero, we also
conclude that the number of second-class constraints is even. Let us check
the consistency of the second-class constraints and the consequences for the
Lagrange multipliers; we start from%
\begin{equation}
\dot{\chi}_{A}=\{\chi _{A},~H_{p}\}\approx h_{A}+u^{\rho }~\{\chi _{A},\chi
_{\rho }\}=h_{A}+u^{\rho }~\Delta _{A\rho }\approx 0.
\end{equation}%
Then, by multiplying with $\Delta ^{BA}$%
\begin{equation}
0\approx \Delta ^{BA}h_{A}+u^{\rho }~\delta _{\rho }^{B}.
\end{equation}%
Therefore, if the index $B$ alludes to a secondary constraint it is%
\begin{equation}
0\approx \Delta ^{BA}h_{A},
\end{equation}%
which should already be a secondary constraint, since we have assumed that
the algorithm is finished (all the secondary constraints have been found).
On the other hand, if the index $B$ alludes to a primary constraint it is%
\begin{equation}
u^{\rho }=-\Delta ^{\rho A}h_{A}.
\end{equation}%
These two results imply that the primary Hamiltonian can be written as
\footnote{%
This Hamiltonian is usually called the \textit{total} Hamiltonian, since it
recognizes the ambiguity associated with the functions $\phi _{a}$ .}%
\begin{equation}
H_{p}=H_{c}+v^{a}~\phi _{a}+h_{A}~\Delta ^{AB}\chi _{B}.  \label{Hprimary}
\end{equation}%
The ambiguity associated with the free functions $v(t)$ implies that only
first-class phase-space functions will unambiguously evolve. For any other
phase space the evolution will be determined modulo \textit{gauge
transformations} generated by the $\phi _{a}$'s. Dirac conjectured that
not only the primary first-class constraints but all the $\gamma _{%
\widetilde{A}}$'s generate gauge transformations. Because of this reason it
is a common practice to use instead the \textit{extended} Hamiltonian%
\begin{equation}
H_{E}=H_{c}+v^{\widetilde{A}}~\gamma _{\widetilde{A}}+h_{A}~\Delta ^{AB}\chi
_{B}
\end{equation}%
without damaging the evolution of the first-class phase-space functions.

The gauge freedom involved in $H_{E}$ can be fully frozen by accompanying
the $\gamma _{\widetilde{A}}$'s with an equal number of independent \textit{%
gauge-fixing conditions }$\xi _{\widetilde{A}}(q,p)\approx 0$.\footnote{%
The conditions $\xi _{\widetilde{A}}(q,p)\approx 0$ must be attainable by
means of gauge transformations generated by the $\gamma _{\widetilde{A}}$'s.}
If the gauge-fixing conditions fulfill $\det \{\gamma _{\widetilde{A}},\xi _{%
\widetilde{B}}\}\not\approx 0$, then the $v^{\widetilde{A}}$'s will be
completely fixed by the requirement that the gauge-fixing conditions must be
consistent with the evolution of the system. Actually $\det \{\gamma _{%
\widetilde{A}},\xi _{\widetilde{B}}\}\not\approx 0$ means that the $\gamma _{%
\widetilde{A}}$'s and the $\xi _{\widetilde{B}}$'s form a second-class set.
In fact no first-class constraint remains since the gauge freedom has been
completely frozen. Not only the gauge-invariant functions --the \textit{%
observables}-- but any phase-space function will so evolve without
ambiguities. Thus the phase space is restricted by the set of conditions $%
\gamma _{\widetilde{A}}\approx 0$, $\xi _{\widetilde{A}}\approx 0$, $\chi
_{A}\approx 0$. Each pair of conditions eliminates one degree of freedom.
Therefore, the d.o.f. are counted by considering the number of pairs of
canonical variables $(q^{n},p_{n})$ and the number of first-class (\textit{%
f.c.}) and second-class (\textit{s.c.}) constraints through the following
formula
\begin{equation}
\text{number of d.o.f.}=\text{number of }(p,q)-\text{number of \textit{f.c.}
constraints}-\dfrac{1}{2}~\text{number of \textit{s.c.} constraints}~.
\label{counting}
\end{equation}%
We will make extensive use of this algorithm in the following subsections.

\subsection{Rotationally pseudoinvariant Lagrangian}

We will propose a toy model that mimics some general features of TEGR
theory, so later we can study its modification, which will possess several
features also present in $f(T)$ gravity. Let us study the following
mechanical Lagrangian~\footnote{%
This model and some of the conclusions drawn here were first 
presented in Ref. \cite{Guzman:2019oth}. However, explicit calculations are given
in the present article.}
\begin{equation}
L=2\left( \frac{d}{dt}\sqrt{z\overline{z}}\right) ^{2}-U(z\overline{z})+\dot{%
z}~\frac{\partial }{\partial z}g(z,\overline{z})+~\dot{\overline{z}}~\frac{%
\partial }{\partial \overline{z}}g(z,\overline{z})~~,  \label{Lrot}
\end{equation}%
where $z$, $\overline{z}$ are complex-conjugate canonical variables. As can
be seen, $L$ is a Lagrangian governing the evolution of a sole dynamical
variable: $z\overline{z}$. In fact the last two terms are just the total
derivative $dg(z,\overline{z})/dt$ and they do not influence the Lagrange
equations. Besides, the first term is a kinetic energy for $\sqrt{z\overline{%
z}}$,
\begin{equation*}
2\left( \frac{d}{dt}\sqrt{z\overline{z}}\right) ^{2}=\frac{1}{2~z\overline{z}%
}~\left( \overline{z}~\dot{z}+z~\dot{\overline{z}}\right) ^{2}~,
\end{equation*}%
and $U$ is a potential for $z\overline{z}$. This means that the Lagrange
equations will govern the evolution of the modulus of the complex variable $%
z $, but the evolution of its phase $z/|z|$ will remain undetermined. We
can notice this fact also at the level of the symmetries of the Lagrangian,
which is \textit{pseudoinvariant} under (\textquotedblleft local\textit{%
\textquotedblright }) time-dependent rotations (it is invariant except for a
total derivative):%
\begin{equation}
z~\longrightarrow ~e^{i\alpha (t)}~z,~~~~\overline{z}~\longrightarrow
~e^{-i\alpha (t)}~\overline{z}~~~~~\ ~~\ \Longrightarrow ~~~~~\ ~~\ \delta L=%
\frac{d}{dt}\delta g(z,\overline{z}).
\end{equation}%
We can recognize some features that resemble the TEGR theory. In fact, the
TEGR Lagrangian is pseudoinvariant under LLT of
the tetrad, so it only governs the dynamics of the metric, but it is unable
to determine the \textquotedblleft orientation\textquotedblright\ of the
tetrad. The analogy is not complete because the boundary term in this toy
model just contains first-order derivatives of the canonical variables,
differing from the case of TEGR on which the boundary term contains second-order derivatives of the tetrad.

Now let us pass to the Hamiltonian formalism, and look for the constraint
algebra. The canonical momenta are defined as
\begin{equation}
p_{z}\equiv \frac{\partial L}{\partial \dot{z}}=\frac{1}{z}~\frac{d}{dt}(z%
\overline{z})+\frac{\partial }{\partial z}g(z,\overline{z}),\ \ \ \ \ p_{%
\overline{z}}\equiv \frac{\partial L}{\partial \dot{\overline{z}}}=\frac{1}{%
\overline{z}}~\frac{d}{dt}(z\overline{z})+\frac{\partial }{\partial
\overline{z}}g(z,\overline{z}),
\end{equation}%
from which it is easily seen that the \textit{primary} constraint is
\begin{equation}
G^{(1)}\equiv ~z\ \left( p_{z}-\frac{\partial g}{\partial z}\right) -%
\overline{z}\ \left( p_{\overline{z}}-\frac{\partial g}{\partial \overline{z}%
}\right) \approx ~0,  \label{G1}
\end{equation}%
which fulfills%
\begin{equation}
\{G^{(1)},z\overline{z}\}~=~0.  \label{observable}
\end{equation}%
In Eq.~\eqref{G1} one recognizes the form of the angular momentum, so $%
G^{(1)}$ is the generator of rotations. Equation \eqref{observable} then
says that $z\overline{z}$ is invariant under rotations. As it happens in any
theory having invariance under rotations, the angular momentum is
conserved;\ however the conservation here is not the result of the dynamical
equations but it appears in the form of a constraint among the canonical
variables. This means that the (conserved) value of the angular momentum
cannot be freely chosen by manipulating the initial conditions; instead the
initial conditions are restricted to satisfy the sole allowed value $%
G^{(1)}=0$. The reason why the angular momentum behaves in such way is
because not only is the Lagrangian (pseudo)invariant under rotations, like
the Lagrangian of a particle in a central potential, but its very
dynamical variable $z\overline{z}$ is already invariant under (even local)
rotations. These are the features characterizing the so-called gauge
systems, i.e. those systems whose Lagrangians do not give dynamics to each
canonical variable, but only govern some combinations of variables, which
can be recognized through their invariance under (local) \textit{gauge}
transformations. Noticeably, in the case under study, the angular momentum $%
G^{(1)}$ has contributions coming from those terms in $L$ that are linear in
$\dot{z}$, $\dot{\overline{z}}$ (the terms we added to $L$ to make it 
pseudoinvariant). As we know, these extra terms do not affect the
fulfillment of the Lorentz algebra in theories of gravity such as TEGR \cite%
{Ferraro:2016wht}; however they are essential to establish the number of
degrees of freedom of the \textquotedblleft deformed\textquotedblright\
theories, as we are going to see in the next subsection.

The canonical Hamiltonian is
\begin{eqnarray}
H &=&\dot{z}~p_{z}+\dot{\overline{z}}~p_{\overline{z}}-L=2~\left( \frac{d}{dt%
}\sqrt{z\overline{z}}\right) ^{2}+U(z\overline{z})  \notag \\
&=&\frac{1}{8~z\overline{z}}~\left[ z~\left( p_{z}-\frac{\partial g}{%
\partial z}\right) \ +\overline{z}\ \left( p_{\overline{z}}-\frac{\partial g%
}{\partial \overline{z}}\right) \right] ^{2}+U(z\overline{z}),  \label{Hrot}
\end{eqnarray}%
while the \textit{primary} Hamiltonian is
\begin{equation}
H_{p}=H+u(t)~G^{(1)},
\end{equation}%
where $u(t)$ is a Lagrange multiplier. The presence of the last term has a
twofold meaning. On the one hand it means that the form of the Hamiltonian
is ambiguous on the constraint surface, since one can rewrite some of the
canonical variables by using the constraint. On the other hand it implies
that only the quantities $\mathcal{O}(z,\overline{z},p_{z},p_{\overline{z}})$
having rotational invariance (i.e., $\{G^{(1)},\mathcal{O}\}\approx 0$) will
unambiguously evolve. The other (non-gauge-invariant) quantities are not
\textit{observables}; their evolution will remain ambiguous as long as the
function $u(t)$ remains unknown.

For consistency reasons, $G^{(1)}$ should evolve without leaving the
constraint surface; i.e. $\dot{G}^{(1)}=\{G^{(1)},H_{p}\}\approx 0$. If this
condition were not fulfilled, then one would impose new (\textit{secondary}%
) constraints to have a consistent evolution. Since $\{G^{(1)},z\overline{z}%
\}$ is zero, then%
\begin{equation}
\{G^{(1)},~H_{p}\}\approx 0~~~~\Longleftrightarrow ~~~~\{G^{(1)},~z\ p_{z}+%
\overline{z}\ p_{\overline{z}}-z\ \frac{\partial g}{\partial z}-\overline{z}%
\ \frac{\partial g}{\partial \overline{z}}\}\approx 0.
\end{equation}%
As can be easily verified $\dot{G}^{(1)}=\{G^{(1)},H_{p}\}=0$ (i.e. it is
zero throughout the phase space, not only on the constraint surface). Therefore,
no \textit{secondary} constraints appear in this example. There is a unique
(necessarily) \textit{first-class} constraint, so one degree of freedom is
removed, and the system is left with just one genuine degree of freedom \cite%
{Sundermeyer1,Sundermeyer2,Henneaux}. As said, $z\overline{z}$ is the gauge
invariant (observable) associated to the unique physical degree of freedom.

The analogies between the toy model and the TEGR theory are summarized in
Table \ref{TM}. Notice that in the table we do not list all TEGR
constraints; the discussion on the number and physical interpretation of
them can be found in Sec. \ref{sec:ftdof}.
\begin{table}[h]
\begin{tabular}{|c||c|c|}
\hline
& toy model & TEGR \\ \hline\hline
coordinates & $z,\,\overline{z}$ & $E^a_{\ \mu}$ \\ \hline
gauge parameter & $\alpha(t)$ & $\Lambda^{a}_{\ b}(x)$ \\ \hline
gauge symmetry & rotations & Lorentz transformations \\ \hline
primary constraint(s) & $G^{(1)}$ & $G^{(1)}_{ab}$ \\ \hline
degrees of freedom & 1 & 2 \\ \hline
observable & $|z|$ & two polarizations of $g_{\mu\nu}$ \\ \hline
\end{tabular}%
\caption{Comparison between the rotationally pseudoinvariant toy model and
TEGR.}
\label{TM}
\end{table}

\subsection{Modified pseudoinvariant rotational Lagrangian}

Let us deform the mechanical toy model given by the Lagrangian \eqref{Lrot} and
replace the pseudoinvariant Lagrangian $L$ with a function of itself:
\begin{equation}
\mathcal{L}=f(L).  \label{f(L)}
\end{equation}%
The theory described by the Lagrangian $\mathcal{L}=f(L)$ is dynamically
equivalent to the one governed by the Jordan-frame Lagrangian that includes
an additional dynamical variable $\phi $:
\begin{equation}
\mathcal{L}=\phi ~L-V(\phi ).  \label{Jordan}
\end{equation}%
In fact, the Lagrange equation for $\phi $ is%
\begin{equation}
L~=~V^{\prime }(\phi ).  \label{eom1}
\end{equation}%
So, the dynamics says that $\mathcal{L}$ in Eq.~\eqref{Jordan} is the
Legendre transform of $V(\phi )$; therefore, $\mathcal{L}$ is a function $%
f(L)$. Each choice of $V$ equals a choice of $f$; the inverse Legendre
transform then implies%
\begin{equation}
\phi =f^{\prime }(L).  \label{fprime}
\end{equation}%
On the other hand, the Lagrange equation \eqref{eom1} also says that the
dynamics of $\phi $ is completely determined by the dynamics of $z(t)$ and $%
\overline{z}(t)$ through the function $L(z(t),\overline{z}(t),~\dot{z}(t),~%
\dot{\overline{z}}(t))$. We remark that, although $L$ comes with a total
derivative, $\mathcal{L}$ is not pseudoinvariant because the total
derivative in Eq.~\eqref{Jordan} is multiplied by $\phi $. So, the system
described by the Lagrangian $\mathcal{L}$ has, in principle, two degrees of
freedom, let us say\ $z$ and $\overline{z}$. Nevertheless, there is a
particular case where the number of degrees of freedom reduces to one:\ when
the function $g$ in Eq.~\eqref{Lrot} has the form $g(z,\overline{z})=v(z%
\overline{z})$. In that case $\mathcal{L}$ in Eq.~\eqref{Jordan} depends
only on $z\overline{z}$ and $\phi $; but, as already said, the dynamics for $%
\phi $ is linked to the one for $z\overline{z}$. This alternative (one or
two degrees of freedom) should be reflected by the Dirac-Bergmann algorithm
for the Hamiltonian formalism of this system.

Let us compute the canonical momenta associated with $\phi $, $z$ and $%
\overline{z}$:
\begin{equation}
G_{\pi }^{(1)}\equiv \pi =\frac{\partial \mathcal{L}}{\partial \dot{\phi}}%
\approx 0,
\end{equation}%
\begin{equation}
p_{z}=\frac{\phi }{z}~\frac{d}{dt}(z\overline{z})+\phi ~\frac{\partial }{%
\partial z}g(z,\overline{z}),\ \ \ \ \ p_{\overline{z}}=\frac{\phi }{%
\overline{z}}~\frac{d}{dt}(z\overline{z})+\phi ~\frac{\partial }{\partial
\overline{z}}g(z,\overline{z}).  \label{momenta}
\end{equation}%
We easily get the angular momentum constraint
\begin{equation}
G^{(1)}\equiv ~z\ \left( p_{z}-\phi ~\frac{\partial g}{\partial z}\right) -%
\overline{z}\ \left( p_{\overline{z}}-\phi ~\frac{\partial g}{\partial
\overline{z}}\right) \approx ~0.  \label{G}
\end{equation}%
Notice that the piece of $G^{(1)}$ which comes from the boundary term in $L$
is now multiplied by $\phi $. In consequence, the dynamical system defined
by Eq.~\eqref{Jordan} has two primary constraints whose Poisson bracket is
\begin{equation}
\{G^{(1)},G_{\pi }^{(1)}\}~=-z\ \frac{\partial g}{\partial z}+\overline{z}\
\frac{\partial g}{\partial \overline{z}}.  \label{secondclass}
\end{equation}%
The canonical Hamiltonian is
\begin{eqnarray}
\mathcal{H}\  &=&\ \dot{z}~p_{z}+\dot{\overline{z}}~p_{\overline{z}}-%
\mathcal{L}=2~\phi ~\left( \frac{d}{dt}\sqrt{z\overline{z}}\right) ^{2}+\phi
~U(z\overline{z})+V(\phi )  \notag \\
&=&\ \frac{1}{8~\phi ~z\overline{z}}~\left[ z\ \left( p_{z}-\phi \ \frac{%
\partial g}{\partial z}\right) +\overline{z}\ \left( p_{\overline{z}}-\phi \
\frac{\partial g}{\partial \overline{z}}\right) \right] ^{2}+\phi ~U(z%
\overline{z})+V(\phi ),  \label{canonicalH}
\end{eqnarray}%
and the primary Hamiltonian is%
\begin{equation}
\mathcal{H}_{p}~=~\mathcal{H}+u^{\pi }(t)~G_{\pi }^{(1)}+u(t)~G^{(1)},
\label{Hamiltonian}
\end{equation}%
where $u^{\pi }$, $u$ are Lagrange multipliers. We must evaluate the
evolution of the primary constraints to look for secondary constraints:%
\begin{eqnarray}
\dot{G}_{\pi }^{(1)} &=&\{G_{\pi }^{(1)},\mathcal{H}_{p}\}=\{\pi ,\mathcal{H}%
_{p}\}  \notag \\
&=&\frac{(z\ p_{z}+\overline{z}\ p_{\overline{z}})^{2}~}{8~\phi ^{2}~z%
\overline{z}}-\frac{1}{8~z\overline{z}}~\left( z\ \frac{\partial g}{\partial
z}+\overline{z}\ \frac{\partial g}{\partial \overline{z}}\right) ^{2}-U(z%
\overline{z})-V^{\prime }(\phi )~+u(t)~\left( z\ \frac{\partial g}{\partial z%
}-\overline{z}\ \frac{\partial g}{\partial \overline{z}}\right) ~  \notag \\
&=&L-V^{\prime }(\phi )+\left( u(t)-\frac{d}{dt}\ln \sqrt{\frac{z}{\overline{%
z}}}\right) \left( z~\frac{\partial g}{\partial z}-\overline{z}~\frac{%
\partial g}{\partial \overline{z}}\right),  \label{Gpidot}
\end{eqnarray}%
\begin{equation}
\dot{G}^{(1)}=\{G^{(1)},\mathcal{H}_{p}\}=u^{\pi }~\{G^{(1)},G_{\pi
}^{(1)}\}=-u^{\pi }~\left( z\ \frac{\partial g}{\partial z}-\overline{z}\
\frac{\partial g}{\partial \overline{z}}\right).  \label{Gdot}
\end{equation}%
Therefore there are three different ways to guarantee the consistency of the
evolution, which we proceed to study in three separate cases.

\subsubsection{Case (i)}

If $g(z,\overline{z})\neq v(z\overline{z})$, i.e. $z\ \frac{\partial g}{%
\partial z}-\overline{z}\ \frac{\partial g}{\partial \overline{z}}\neq 0$,
then we guarantee the consistency of the evolution by choosing the Lagrange
multipliers in the following way:
\begin{equation}
u^{\pi }~=~0~,~~~~~~~~~~~~~~u(t)=-\frac{L(t)-V^{\prime }(\phi (t))}{z(t)~%
\frac{\partial g}{\partial z}(t)-\overline{z}(t)~\frac{\partial g}{\partial
\overline{z}}(t)}~+\frac{d}{dt}\ln \sqrt{\frac{z(t)}{\overline{z}(t)}}.
\label{multipliers}
\end{equation}%
The system has no secondary constraints; the only constraints $G^{(1)}$, $%
G_{\pi }^{(1)}$ are second class, since $\{G^{(1)},G_{\pi }^{(1)}\}$ in Eq.~%
\eqref{secondclass} is different from zero. So, they remove only one degree
of freedom \cite{Sundermeyer1,Sundermeyer2,Henneaux}; there are two genuine
degrees of freedom among the variables $(z,~\overline{z},~\phi )$. Notice
that, differing from $f(T)$ gravity, no gauge freedom is left in this system
since both Lagrange multipliers have been fixed, so the primary Hamiltonian
completely determines the evolution of the variables. In particular, the
evolution of $\phi $ is given by the equation%
\begin{equation}
\dot{\phi}=\{\phi ,\mathcal{H}_{p}\}=u^{\pi }=0,  \label{dotphi}
\end{equation}%
which means that $\phi $ is a free constant. The equation for $z(t)$ is%
\begin{equation}
\dot{z}=\{z,\mathcal{H}_{p}\}=\frac{1}{4~\phi ~\overline{z}}~\left( z\ p_{z}+%
\overline{z}\ p_{\overline{z}}-\phi ~z\ \frac{\partial g}{\partial z}-\phi ~%
\overline{z}\ \frac{\partial g}{\partial \overline{z}}\right) +u(t)~z=\dot{z}%
-z~\frac{L-V^{\prime }(\phi )}{z~\frac{\partial g}{\partial z}-\overline{z}~%
\frac{\partial g}{\partial \overline{z}}},
\end{equation}%
so the Lagrange equation $L-V^{\prime }(\phi )=0$ [see
Eq.~\eqref{eom1}] is
obtained. The Lagrange multiplier $u$ then becomes%
\begin{equation}
u(t)=\frac{d}{dt}\ln \sqrt{z(t)/\overline{z}(t)}.
\end{equation}
By combining Eqs.~\eqref{eom1} and \eqref{dotphi} one gets%
\begin{equation}
L=\text{const \ \ \ \ \ and \ \ \ \ \ }\mathcal{L}=\text{const.}
\end{equation}%
Instead of $\dot{p}_{z}$, let us compute the evolution of the
rotationally invariant quantity $z\ p_{z}-\phi ~z\ \partial g/\partial z$:
\begin{eqnarray}
\frac{d}{dt}\left( z\ p_{z}-\phi ~z\ \frac{\partial g}{\partial z}\right)
&=&\{z\ p_{z}-\phi ~z\ \frac{\partial g}{\partial z},\mathcal{H}_{p}\}=\{z\
p_{z}-\phi ~z\ \frac{\partial g}{\partial z},\mathcal{H}\}  \notag \\
&=&\frac{1}{8~\phi ~z\overline{z}}~\left( z\ p_{z}+\overline{z}\ p_{%
\overline{z}}-\phi ~z\ \frac{\partial g}{\partial z}-\phi ~\overline{z}\
\frac{\partial g}{\partial \overline{z}}\right) ^{2}-\phi ~z~\frac{\partial
U(z\overline{z})}{\partial z}  \notag \\
&=&\mathcal{H}-\phi ~U(z\overline{z})-V(\phi )-\phi ~z\overline{z}~U^{\prime
}(z\overline{z}).  \label{motioneq0}
\end{eqnarray}%
By replacing this with Eqs.~\eqref{momenta}, \eqref{canonicalH} and \eqref{dotphi}%
, one obtains%
\begin{equation}
\frac{d^{2}}{dt^{2}}(z\overline{z})=2~\left( \frac{d}{dt}\sqrt{z\overline{z}}%
\right) ^{2}-~z\overline{z}~U^{\prime }(z\overline{z}),  \label{motioneq}
\end{equation}%
or%
\begin{equation}
4~\frac{d^{2}}{dt^{2}}\sqrt{z\overline{z}}=-~\frac{dU}{d\sqrt{z\overline{z}}},  \label{motioneq2}
\end{equation}%
which amounts to the conservation of $h=2\left( d(\sqrt{z\overline{z}}%
)/dt\right) ^{2}+U$. Equation \eqref{motioneq} coincides with the Lagrange
equation for the system described by the Lagrangian $L$.\footnote{%
Analogously, in $f(T)$ gravity the solutions with constant $T$
satisfy the Einstein equations (although the cosmological and
gravitational constants are shifted).} However the Lagrangian
$\mathcal{L}$ still makes a difference with $L$. This is because the
dynamics governed by $\mathcal{L}$ imposes a new constant of motion
besides $h$: $L$ has to be a constant as well. While
Eq.~\eqref{motioneq} only fixes the evolution of $|z(t)|$, the
condition $L=$ const is an additional requirement that involves
the phase of $z(t)$ in
the total derivative term of $L$.\footnote{%
In $f(T)$ gravity, it involves the orientation of the tetrad which affects
the boundary term of the TEGR Lagrangian.} Therefore, differing from $L$,
the Lagrangian $\mathcal{L}$ governs the evolutions of both the modulus and
the phase of $z$ (notice, however, that the initial phase is irrelevant due
to the \textit{global} rotational symmetry).

In sum, the system described by the Lagrangian \eqref{Jordan} has two
degrees of freedom:\ one of them is $|z|^{2}=z\overline{z}$ whose dynamics
does not differ from the one described by the Lagrangian \eqref{Lrot}; in
both cases we arrive at the conserved quantity $h=2\left( d(\sqrt{z\overline{%
z}})/dt\right) ^{2}+U$. Once the evolution of $|z|$ is determined by
the choices of the initial value $|z(t_{o})|$ and the constant of
motion $h$, the evolution of the phase of $z$, which is the
remaining degree of freedom, is determined by the condition $L(t)=$
const. The value of this constant connects with the value of
$\phi $ through Eq.~\eqref{eom1}. There is no other physics
associated with $\phi $, over and above the one related to the phase
of $z$. In the analogy with $f(T)$ gravity, $\phi $ could then be 
regarded as a variable carrying information about the
\textquotedblleft orientation\textquotedblright\ of the tetrad,
which would be partially determined by the dynamical equations. We
will discuss this issue more later.

\subsubsection{Case (ii)}

If $g(z,\overline{z})=v(z\overline{z})$, i.e. $z\ \frac{\partial g}{\partial
z}-\overline{z}\ \frac{\partial g}{\partial \overline{z}}=0$, then $\mathcal{%
L}$ depends only on $|z|^{2}=z\overline{z}$ and $\phi $. The constraints $%
G^{(1)}$,$~G_{\pi }^{(1)}$ commute [see Eq.~\eqref{secondclass}],
and the Lagrange multipliers $u$, $u^{\pi }$ are not determined by
the Eqs.~\eqref{Gpidot} -- \eqref{Gdot}. While $\dot{G}^{(1)}$ is
zero, the consistency
of $G_{\pi }^{(1)}$ leads to the \textit{secondary} constraint%
\begin{equation}
G^{(2)}=L-V^{\prime }(\phi )\approx 0,  \label{G2}
\end{equation}%
which recovers Eq.~\eqref{eom1}, and tells that the values of $\phi $
are now linked to those of $z\overline{z}$.

If $g(z,\overline{z})=v(z\overline{z})$, then $L$ is invariant under
rotations. Therefore%
\begin{equation}
\{G^{(1)},G^{(2)}\}=0.
\end{equation}%
Besides%
\begin{equation}
\{G_{\pi }^{(1)},G^{(2)}\}=V^{\prime \prime }(\phi ).
\end{equation}%
Let us examine the consistency of $G^{(2)}$ under the time evolution of the
system:%
\begin{equation}
\dot{G}^{(2)}=\{G^{(2)},\mathcal{H}_{p}\}=\{G^{(2)},\mathcal{H}%
\}+u~\{G^{(2)},G^{(1)}\}+u^{\pi }~\{G^{(2)},G_{\pi }^{(1)}\}=\{L,\mathcal{H}%
\}-u^{\pi }~V^{\prime \prime }(\phi ).  \label{consistence2}
\end{equation}%
If $V^{\prime \prime }(\phi )\neq 0$, then the consistency can be guaranteed
by choosing $u^{\pi }$,\footnote{%
In the Legendre transform it is $V^{\prime \prime }(\phi )^{-1}=f^{\prime
\prime }(L)$.}
\begin{equation}
u^{\pi }(t)=V^{\prime \prime }(\phi )^{-1}~\{L,\mathcal{H}\}=V^{\prime
\prime }(\phi )^{-1}~\frac{dL}{dt}.
\end{equation}%
In such a case we are left with a first-class constraint $G^{(1)}$ and two
second-class constraints $G_{\pi }^{(1)}$, $G^{(2)}$. So, two degrees of
freedom are suppressed by the constraint structure. Since we started with
three dynamical variables, $z$, $\overline{z}$ and $\phi $, the system
has one genuine degree of freedom. The observable (gauge-invariant) variable
is $z\overline{z}$. The phase of $z$ remains as a gauge freedom; it is not
determined by the evolution since the Lagrange multiplier $u(t)$ has not
been fixed.

The dynamical equation for $\phi $,%
\begin{equation}
\dot{\phi}=\{\phi ,\mathcal{H}_{p}\}=u^{\pi }(t)=V^{\prime \prime }(\phi
)^{-1}~\frac{dL}{dt}=f^{\prime \prime }(L)~\frac{dL}{dt}=\frac{d}{dt}%
f^{\prime }(L),
\end{equation}%
does not contain new information since it can also be obtained by
differentiating Eq.~\eqref{G2}; in particular, it does not constrain $L$ to
be a constant. The evolution of $\phi $ is then entirely determined by the
evolution of $z\overline{z}$ through Eq.~\eqref{G2}. On the other hand, the
evolution of \ $z\overline{z}$ will be different than in case (i); this
is because $\phi $ is no longer a constant [as it is in case (i)]. This
does not mean that $\phi $ does not have a role; the reader must remember
that $\phi $ exists because the Lagrangian $L$ has been replaced with $%
\mathcal{L}=f(L)$. In fact, the lhs of Eq.~\eqref{motioneq0}, which is
the equation we used to obtain the evolution of $z\overline{z}$, will now
generate an additional term associated with $\dot{\phi}$. Since $z\
p_{z}-\phi ~z\ \partial g/\partial z=\phi ~d(z\overline{z})/dt$ [see Eq.~%
\eqref{momenta}], the new term will be $\dot{\phi}~d(z\overline{z})/dt$.
Thus, the dynamical equation \eqref{motioneq} for $z\overline{z}$ will now
read%
\begin{equation}
\phi ^{-1}~\dot{\phi}~\frac{d}{dt}(z\overline{z})+\frac{d^{2}}{dt^{2}}(z%
\overline{z})=2~\left( \frac{d}{dt}\sqrt{z\overline{z}}\right) ^{2}-~z%
\overline{z}~U^{\prime }(z\overline{z}),
\end{equation}%
or%
\begin{equation}
4~\frac{d}{dt}\ln f^{\prime }(L)~\frac{d}{dt}(\sqrt{z\overline{z}})+4~\frac{%
d^{2}}{dt^{2}}\sqrt{z\overline{z}}=-~\frac{dU}{d\sqrt{z\overline{z}}}.
\end{equation}%
This is the result we were expecting, because it is the Lagrange equation
for a Lagrangian $\mathcal{L}=f(L)$ that depends exclusively on $z\overline{z%
}$.

\subsubsection{Case (iii)}

As can be seen in Eqs.~\eqref{Gpidot} and \eqref{Gdot}, the
consistency of the evolutions of both primary constraints $G_{\pi
}^{(1)}$ and $G^{(1)}$ are affected by the quantity $z\ \partial
g/\partial z-\overline{z}\
\partial g/\partial \overline{z}$. This quantity vanishes if $g(z,\overline{z%
})=v(z\overline{z})$, as considered in case (ii), which means that $%
\mathcal{L}=f(L)$ becomes invariant under local rotations, and the system is
left with only one degree of freedom. However, we could still consider
another possibility: the condition
\begin{equation}
z\ \frac{\partial g}{\partial z}-\overline{z}\ \frac{\partial g}{\partial
\overline{z}}=0  \label{local}
\end{equation}%
is satisfied only in some region of the constraint surface. For instance,
let us consider a function $g(z,\overline{z})=g(z+\overline{z})$; then
\begin{equation}
g(z,\overline{z})=g(z+\overline{z})~~~~\Longrightarrow ~~~~z\dfrac{\partial g%
}{\partial z}-\overline{z}\dfrac{\partial g}{\partial \overline{z}}=(z-%
\overline{z})\,g^{\prime }.
\end{equation}%
We see that the relevant quantity for our analysis vanishes if $z$ is real.
Therefore the real solutions, if they exist, would work as in case (ii). Since
the phase of $z$ has been frozen to be zero, no extra d.o.f. would be left
in these solutions.

The condition \eqref{local} defines a hypersurface in the phase space. The
intersection of this hypersurface with the constraint surface, if it exists,
would constitute a subspace where the degree of freedom associated with the
phase of $z$ does not manifest itself, since the Lagrangian $\mathcal{L}=f(L)
$ would turn out to be invariant under\textit{\ infinitesimal} local rotations $%
\delta z=i~\alpha (t)~z$:%
\begin{equation}
\delta \mathcal{L}=\delta f(L)=f^{\prime }(L)~\delta L=f^{\prime }(L)~\frac{d%
}{dt}\delta g=f^{\prime }(L)~\frac{d}{dt}\left[ \delta z\ \frac{\partial g}{%
\partial z}+\delta \overline{z}\ \frac{\partial g}{\partial \overline{z}}%
\right] =i~f^{\prime }(L)~\frac{d}{dt}\left[ \alpha (t)~\left( z\ \frac{%
\partial g}{\partial z}-\overline{z}\ \frac{\partial g}{\partial \overline{z}%
}\right) \right] =0.
\end{equation}%
Thus, we should wonder about the existence of solutions to the
equations of motion lying on the subspace defined by
Eq.~\eqref{local} and the constraints. These solutions should not
contain a d.o.f. associated with the phase of $z$; they would remain
as solutions to the equations of motion under infinitesimal local
rotations. These solutions would evidence just one degree of
freedom: the one related to the modulus of $z$. Therefore, the
Lagrangian $\mathcal{L}=f(L)$ could lead to solutions displaying one
or two degrees of freedom, depending on which region of the
constraint surface they occupy [i.e., depending on whether they satisfy
the condition \eqref{local} or not].

\bigskip

In sum, in the Jordan frame we rewrite the Lagrangian $\mathcal{L}=f(L)$ as $%
\mathcal{L}=\phi ~L-V(\phi )$. If the boundary term $\dot{g}(z,\overline{z})$
present in $L$ is such that $z\ \partial g/\partial z-\overline{z}\ \partial
g/\partial \overline{z}\neq 0$, then an extra degree of freedom associated
with the phase of $z$ will manifest itself. In the Jordan frame, the extra
degree of freedom comes from the free choice of the \textit{constant} $\phi $
which, on its side, determines the phase of $z$ through the condition $%
L=V^{\prime }(\phi )=$ const. Instead, if $z\ \partial g/\partial z-%
\overline{z}\ \partial g/\partial \overline{z}=0$, then $\mathcal{L}=f(L)$
will not be sensitive to the phase of $z$, so $\phi $ cannot be associated
with an extra degree of freedom but will be entirely determined by $z%
\overline{z}$ through the equation $G^{(2)}=L-V^{\prime }(\phi )\approx 0$
without imposing any condition on the value of $L$. However the fact that $%
\phi $ is not constrained to be a constant will imply an additional term in
the dynamical equation for the modulus of $z$, as can be straightforwardly
verified in the Lagrange equations for the Lagrangian $\mathcal{L}=f(L)$.
Besides, if there were solutions such that $z\ \partial g/\partial z-%
\overline{z}\ \partial g/\partial \overline{z}$ cancels out, then these
solutions will remain as solutions of the equations of motion under
infinitesimal local perturbations of the phase of $z$; therefore they would
just exhibit the degree of freedom associated with $z\overline{z}$.

\bigskip

\section{$f(T)$ gravity: a modified Lorentzian pseudoinvariant Lagrangian}

\label{sec:ftdof}

\subsection{Summary of d.o.f. counting in $f(T)$ gravity}

The modified rotationally pseudoinvariant system of Sec. \ref%
{sec:toymodel} is useful to understand several features of $f(T)$
gravity, since the latter consists in the modification of the
Lorentzian pseudoinvariant TEGR Lagrangian. Due to the inherent
complications of the dynamical equations of $f(T)$ gravity, the
Jordan-frame formalism has been used for the analysis of the
constraint algebra and the counting of d.o.f.
\cite{Li:2011rn,Ferraro:2018tpu}. Reference \cite{Li:2011rn} used the
first-order Hamiltonian formalism developed in Refs. 
\cite{Maluf:2001rg,daRochaNeto:2011ir} as a base for computing the
constraint structure of $f(T)$ gravity. Instead, Ref.
\cite{Ferraro:2018tpu} used the canonical Hamiltonian formalism for
TEGR described in Ref. \cite{Ferraro:2016wht}.\footnote{%
In Ref.~\cite{Ferraro:2016wht} the TEGR Lagrangian was expressed in the form $%
L_{TEGR}~=~E~T~=~(1/2)~E~\partial _{\mu }E_{\,\,\nu }^{g}~\partial
_{\rho }E_{\,\,\lambda }^{h}~e_{c}^{\mu }e_{e}^{\nu }e_{d}^{\rho
}e_{f}^{\lambda }~M_{gh}^{\ \,\,\,\,cedf}$, where $M_{gh}^{\
\,\,\,cedf}\doteq 2~\eta _{gh}\eta ^{c[d}\eta ^{f]e}-4~\delta
_{g}^{[d}\eta ^{f][c}\delta _{h}^{e]}+8~\delta _{g}^{[c}\eta
^{e][d}\delta _{h}^{f]}$ is the so-called supermetric.} While in
Ref.~\cite{Li:2011rn} the authors claimed that $f(T)$ gravity has
$n-1$ extra d.o.f. in dimension $n$, the outcome of the counting of
d.o.f. in Ref. \cite{Ferraro:2018tpu} gave only one extra d.o.f. in
arbitrary dimension. More evidence that speaks in favor of only one
d.o.f. can be found in Ref.~\cite{Ferraro:2018axk}, where the extra
d.o.f. was identified with a scalar field which partially determines
the orientation of the tetrad field. Other classes of modified
teleparallel gravities might have a different number of d.o.f.
\cite{Blixt:2018znp,Blixt:2019mkt}.

In what follows we will summarize some key findings that are essential for
the understanding of the counting of degrees of freedom in $f(T)$ gravity.
The notation in what comes next will be borrowed from Ref. \cite{Ferraro:2018tpu}%
; the reader can find all the definitions and details there. The constraints
of $f(T)$ gravity can be counted and classified as follows:

\begin{itemize}
\item One primary constraint $G^{(1)}_{\pi}$ coming from the vanishing of
the momentum conjugate to the auxiliary scalar field $\phi$.

\item $n$ primary constraints $G^{(1)}_a$ coming from the absence of $%
\partial_0 E^c_0$ in the Lagrangian (analogous to electromagnetism).

\item $n(n-1)/2$ primary constraints $G^{(1)}_{ab}$ associated with Lorentz
invariance (also appearing in TEGR).

\item $n$ secondary constraints $G^{(2)}_{\mu}$ due to the diffeomorphism
invariance (same constraints as in GR).
\end{itemize}

From the whole set of primary and secondary constraints of the theory, there
are only two nonvanishing Poisson brackets. These correspond to
\begin{equation}
\{G_{ab}^{(1)}(\mathbf{x}),G_{\pi }^{(1)}(\mathbf{y})\}\approx F_{ab}~\delta
(\mathbf{x}-\mathbf{y}),  \label{pblorentz}
\end{equation}%
and
\begin{equation}
\{G_{0}^{(2)}(\mathbf{x}),G_{\pi }^{(1)}(\mathbf{y})\}\approx F_{\phi
}~\delta (\mathbf{x}-\mathbf{y}),  \label{pbscalar}
\end{equation}%
where $F_{ab}$, $F_{\phi }$ are
\begin{equation}
F_{ab}=4E~\partial _{i}E_{j}^{c}~e_{[b}^{0}e_{a}^{i}e_{c]}^{j},\ \ \ \ \ \
\ F_{\phi }=E\left( T-V^{\prime }(\phi )\right) .  \label{eFs}
\end{equation}%
The functions $F_{ab}$, $F_{\phi }$ are key in determining the number of
physical d.o.f. of the theory. They enter the matrix of Poisson brackets
$C_{\hat{\rho}\rho }$, so they determine the rank of $C_{\hat{\rho}\rho }$
and the separation of the constraints into first and second class. These
functions can be arranged to compose a vector $\mathbf{F}$,
\begin{equation}
\mathbf{F}=(F_{\phi },F_{01},F_{02},\ldots ,F_{(n-2)(n-1)})\equiv
(F_{0},F_{1},F_{2},\ldots ,F_{n(n-1)/2}).
\end{equation}%
We also define the vector $\mathbf{G}$,%
\begin{equation}
\mathbf{G}=(G_{0}^{(2)},G_{01}^{(1)},G_{02}^{(1)},\ldots
,G_{(n-2)(n-1)}^{(1)})\equiv (G_{0},G_{1},G_{2},\ldots ,G_{n(n-1)/2}),
\end{equation}%
to write the brackets \eqref{pblorentz}-\eqref{pbscalar} in a vector form:%
\begin{equation}
\{\mathbf{G}(\mathbf{x}),G_{\pi }^{(1)}(\mathbf{y})\}\approx \mathbf{F}%
~\delta (\mathbf{x}-\mathbf{y}).
\end{equation}%
This vector equation can be \textquotedblleft rotated\textquotedblright\ to
have all the components of $\mathbf{F}$ but one equal to zero. In other
words, the constraints $G_{ab}^{(1)}$ and $G_{0}^{(2)}$ can be rearranged
by linearly combining them to have all the brackets \eqref{pblorentz}--\eqref{pbscalar} 
but one equal to zero. Therefore, the brackets %
\eqref{pblorentz}--\eqref{pbscalar} just mean that one combination of $%
G_{ab}^{(1)}$'s and $G_{0}^{(2)}$ will fail to be a first-class constraint.
That combination together with $G_{\pi }^{(1)}$ will make up a (unique) pair
of second-class constraints. As is known, the \textit{pairs} of second-class
constraints count as \textit{individual} first-class constraints in the
counting of d.o.f. \eqref{counting}. So, although $f(T)$ gravity in the
Jordan frame has an additional constraint $G_{\pi }^{(1)}$ compared with
TEGR, the number to be subtracted in the counting of d.o.f. \eqref{counting} 
will not change because one of the first-class constraints
of TEGR has joined $G_{\pi }^{(1)}$ to make up a pair of second-class
constraints. Since $f(T)$ gravity in the Jordan frame has an extra pair of
canonical variables $(\phi ,\pi )$, one concludes that $f(T)$ gravity
contains an extra d.o.f. irrespective of the dimension of the spacetime.
The extra d.o.f. is the other side of the coin of the reduction of the gauge
freedom, since a combination of Lorentz constraints now takes part in a
second-class constraint; thus, it stops generating a Lorentz gauge
transformation. Therefore, the orientation of the tetrad in $f(T)$ gravity
would be partially determined through the choice of the extra d.o.f. in the
initial conditions. Which combination of Lorentz constraints no
longer generates a gauge transformation will depend on the value of $\mathbf{%
F}$ for each solution; we just mention that $F_{\phi }$ will be
dynamically zero [cf. Eq.~\eqref{eom1}].

\subsection{Lessons of the toy model for $f(T)$ gravity}

Concerning the comparison between the toy model and $f(T)$ gravity,
we see that the Poisson brackets \eqref{pblorentz} are analogous to
its toy model counterpart $\{G^{(1)},G_{\pi }^{(1)}\}$ defined in
Eq.~\eqref{secondclass}. The analogy implies that the functions
$F_{ab}$ somehow play a role analogous to $z\ \partial g/\partial
z-\overline{z}\ \partial g/\partial
\overline{z}$.\footnote{%
The toy model does not have an analogue for the bracket \eqref{pbscalar}
because it is not invariant under reparametrizations. The
reparametrization invariance can be considered by adding a
\textit{lapse} function $N(t)$ to the set of dynamical variables
(see for instance Ref.~\cite{Ferraro2016}).} While $G^{(1)}$
corresponds to the rotational gauge symmetry of the toy model,
$G_{ab}^{(1)}$ is related to the Lorentz gauge symmetry of TEGR. Both
symmetries will be lost in the modified models, due to the
nonvanishing of the brackets in Eqs.~\eqref{secondclass} and
\eqref{pblorentz}, respectively. However, in $f(T)$ there is still
room for a subset of Lorentz transformations that keep being a
symmetry of the theory; this subset is determined by the value of
the vector $\mathbf{F}$ in each solution.

The analogy between $z\ \partial g/\partial z-\overline{z}\ \partial
g/\partial \overline{z}$ and $F_{ab}$ also appears at the Lagrangian level
in the analysis of the pseudoinvariance of $L$ and $L_{\text{TEGR}}=E\ T$. In
fact, the change of $L$ under an infinitesimal rotation of angle $\alpha (t)$
[i.e., $\delta z=i~\alpha (t)~z$],
\begin{equation}
~\delta L=\delta \frac{dg}{dt}=\frac{d}{dt}\delta g=\frac{d}{dt}\left[
i~\alpha (t)\left( z\ \frac{\partial g}{\partial z}-\overline{z}\ \frac{%
\partial g}{\partial \overline{z}}\right) \right],  \label{Lvar}
\end{equation}%
is governed by $z\ \partial g/\partial z-\overline{z}\ \partial g/\partial
\overline{z}$. On the other hand, the infinitesimal Lorentz transformation
of $L_{\text{TEGR}}$ can be obtained from the expression \eqref{bterm}, rewritten
as $L_{\text{TEGR}}=-ER+2\partial _{\mu }(ET^{\mu })$. In varying it, we must take
into account that $ER$ is locally invariant under Lorentz transformations of
the tetrad: it depends exclusively on the metric. Then, the variation of $%
L_{\text{TEGR}}$ is equal to the variation of the boundary term $2~\partial _{\mu
}(ET^{\mu })$, that is
\begin{equation}
\delta L_{TEGR}=2~\delta \partial _{\mu }(ET^{\mu }).  \label{pseudo}
\end{equation}%
Since $T^{\mu }$ is invariant only under global Lorentz
transformations of the tetrad field, Eq.~\eqref{pseudo}
exhibits the pseudo-gauge-invariance of $L_{\text{TEGR}}$. Let us consider
an infinitesimal local Lorentz transformation of the tetrad in the
$a-b$ plane,
\begin{equation}
\delta _{ab}\mathbf{E}^{g}=-\alpha (t,\mathbf{x})~\delta _{\lbrack
a}^{g}~\eta _{b]h}~\mathbf{E}^{h}.  \label{lttetrad}
\end{equation}%
Then the change of $\partial _{\mu }(E~T^{\mu })$ is
\begin{eqnarray}
\delta _{ab}\partial _{\mu }(E~T^{\mu }) &=&\partial _{\mu }(E~\delta
_{ab}T^{\mu })=-\partial _{\mu }[E~g^{\mu \nu }e_{g}^{\rho }~(\delta
_{\lbrack a}^{g}~\eta _{b]h}~E_{\rho }^{h}~\partial _{\nu }\alpha -\delta
_{\lbrack a}^{g}~\eta _{b]h}~E_{\nu }^{h}~\partial _{\rho }\alpha )]  \notag
\\
&=&\partial _{\mu }\left( E~g^{\mu \nu }e_{[a}^{\rho }~\eta _{b]h}~E_{\nu
}^{h}~\partial _{\rho }\alpha \right) =\partial _{\mu }\left( E~e_{[a}^{\rho
}~e_{b]}^{\mu }~\partial _{\rho }\alpha \right) =\partial _{\rho }\alpha
~\partial _{\mu }\left( E~e_{[a}^{\rho }~e_{b]}^{\mu }\right).
\label{Lorentzvar}
\end{eqnarray}%
In this calculation we have only kept terms involving derivatives of the
parameter $\alpha $, because we already know that $L_{\text{TEGR}}$ is not
sensitive to global Lorentz transformations [represented by $\alpha =$const 
in \eqref{lttetrad}].\footnote{Instead, $L$ in the toy model changes even if $\alpha $ is a
constant, as seen in Eq.~\eqref{Lvar}.} Using the standard formulas
$\partial _{\mu }E=E~e_{a}^{\nu }~\partial _{\mu }E_{\nu }^{a}$,
$\partial _{\mu }e_{b}^{\nu }=-e_{a}^{\nu }~e_{b}^{\lambda
}~\partial _{\mu }E_{\lambda }^{a}$, it is possible to show that
\begin{equation}
\partial _{\mu }(E~e_{[a}^{\rho }e_{b]}^{\mu })=3~E~\partial _{\mu
}E_{\lambda }^{c}~e_{[a}^{\rho }~e_{b}^{\mu }~e_{c]}^{\lambda }.
\label{Lvar3}
\end{equation}%
Comparing with Eq.~\eqref{eFs} we see that $\partial _{\mu
}(E~e_{[a}^{0}~e_{b]}^{\mu })=-(3/4)F_{ab}$ . Thus the variation %
\eqref{Lorentzvar} implies that
\begin{equation}
\delta _{ab}L_{TEGR}=-\dfrac{3}{2}~\partial _{0}\alpha ~F_{ab}+2~\partial
_{i}\alpha ~\partial _{\mu }\left( E~e_{[a}^{i}~e_{b]}^{\mu }\right).
\label{LvarF}
\end{equation}%
As seen, both $F_{ab}$ and $z\ \partial g/\partial z-\overline{z}\ \partial
g/\partial \overline{z}$ play a role in the pseudoinvariance of $L_{\text{TEGR}}$
and $L$ respectively.

The modified toy model has shown us that two types of solutions can
exist when the original system possesses pseudoinvariance: the
case-(i) solutions where the Lagrangian is a constant to be chosen
in the initial conditions (it is the extra d.o.f.), and the
case-(iii) solutions where the extra d.o.f. does not manifest itself
since it is subject to satisfying the condition $z\ \partial
g/\partial z-\overline{z}\ \partial g/\partial \overline{z}=0$.
According to Eq.~\eqref{Lvar}, the case-(iii) solutions are made
of points of the configuration space where the Lagrangian is
invariant rather than pseudoinvariant. Analogously, the case-(iii)
solutions of $f(T)$ gravity do not exhibit the extra d.o.f. because
it is subject to canceling out the $F_{ab}$'s. According to Eq.~\eqref{LvarF}, 
the case-(iii) solutions of $f(T)$ gravity are
made of configurations such that $L_{\text{TEGR}}$ is invariant, rather
than pseudoinvariant, under Lorentz transformations depending only
on time [if $\alpha =\alpha (t)$, then $\delta
_{ab}L_{TEGR}=-(3/2)~\dot{\alpha}~F_{ab}$].

The interest in case-(iii) solutions comes from the fact that they give new
dynamics to the original gauge-invariant variables; in $f(T)$ gravity, they
are apt to study modified gravity. In a case-(i) solution, instead, the
dynamics for the components of the metric tensor is the same as in TEGR,
except for the shift of the cosmological and Newton constants due to the
fact that the determinant $E$ is not encapsulated in the function $f$.

The previously remarked analogies between the modified toy model and $f(T)$
gravity are summarized in Table \ref{tmft}.

\begin{table}[h!]
\begin{tabular}{|c||c|c|}
\hline
& modified toy model & $f(T)$ gravity \\ \hline\hline
boundary term & $\dot{z} \frac{dg}{dz} + \dot{\overline{z}}\frac{dg}{d%
\overline{z}}$ & $\partial_{\mu} ( E T^\mu)$ \\ \hline
Poisson bracket & $\{G^{(1)},G^{(1)}_{\pi} \} = -z \frac{dg}{dz} + \overline{%
z}\frac{dg}{d\overline{z}} $ & $\{G^{(1)}_{ab},G^{(1)}_{\pi} \} = F_{ab} $
\\ \hline
lost gauge symmetry & rotation in the plane $(z,\overline{z})$ & a linear
combination of Lorentz transformations \\ \hline
degrees of freedom & $|z| + $ scalar field & two polarizations of $g_{\mu\nu}
+ $ scalar field \\ \hline
\end{tabular}%
\caption{Comparison between the modified rotationally pseudoinvariant toy
model and $f(T)$ gravity}
\label{tmft}
\end{table}

\subsection{Both types of solutions in flat FLRW cosmology}

We will exemplify case-(i) and case-(iii) solutions in $f(T)$ gravity by
revisiting cosmological solutions already existing in the literature, in the
context of flat FLRW cosmology. The commonly used solution is the diagonal
tetrad in Eq. \eqref{cosmotetrad1} with $T=-6H^{2}$ \cite%
{Ferraro:2006jd,Bengochea:2008gz}. It is easy to prove that all the
coefficients $F_{ab}$ are zero, since the tetrad \eqref{cosmotetrad1}
depends only on time, and $F_{ab}$ just involve spatial derivatives. This is
a case-(iii) solution; therefore, the extra d.o.f. does not manifest itself.

On the other hand, the tetrad \eqref{mccosmo} is a case-(i) solution. In
fact, $T$ is a constant $T_{o}$; besides some of the $F_{ab}$'s are
different from zero, namely
\begin{equation}
F_{01}=-\dfrac{4}{3}~r~a(t)^{2}\sin \theta ~,~~~~~~~F_{02}=-\dfrac{2}{3}%
~r~a(t)^{2}\cos \theta ~\cosh \lambda ~,~~~~~~~F_{12}=\dfrac{2}{3}%
~r~a(t)^{2}\cos \theta ~\sinh \lambda  \label{Fab}
\end{equation}%
(the rest of the antisymmetric components $F_{ab}$ are zero on-shell). By
replacing the tetrad \eqref{mccosmo} in the equations of motion of $f(T)$
gravity, one obtains that the scale factor $a(t)$ fulfills the FLRW
equations of general relativity with shifted gravitational and cosmological
constants, as shown in Eq.~\eqref{FRWeom}. The extra d.o.f. $\phi =f^{\prime
}(T_{o})$ is represented by $T_{o}$, which takes part in the tetrad field
through the function $\lambda (t)$ in the same way that $\phi $ enters the
phase of $z$ in the case-(i) solutions of the toy model. Equation \eqref{Fab}
suggests combining the constraints $G_{01}^{(1)}$, $G_{02}^{(1)}$ and $%
G_{12}^{(1)}$ as
\begin{eqnarray}
\mathcal{G}_{01}^{(1)} &=&G_{01}^{(1)},  \notag \\
\mathcal{G}_{02}^{(1)} &=&\sinh \lambda ~G_{02}^{(1)}+\cosh \lambda
~G_{12}^{(1)},  \notag \\
\mathcal{G}_{12}^{(1)} &=&\cos \theta ~G_{01}^{(1)}-2~\sin \theta ~(\cosh
\lambda ~G_{02}^{(1)}+\sinh \lambda ~G_{12}^{(1)}),
\end{eqnarray}%
to get on-shell%
\begin{equation}
\{\mathcal{G}_{01}^{(1)}(\mathbf{x}),G_{\pi}^{(1)}(\mathbf{y})\}\approx
F_{01}~\delta (\mathbf{x}-\mathbf{y})~,~~~~\ ~~~~\ \{\mathcal{G}_{02}^{(1)}(%
\mathbf{x}), G_{\pi }^{(1)}(\mathbf{y})\}\approx 0~,~~~~\ ~~~~\ \{\mathcal{G}%
_{12}^{(1)}(\mathbf{x}),G_{\pi }^{(1)}(\mathbf{y})\}\approx 0.
\end{equation}%
Therefore, $\mathcal{G}_{01}^{(1)}$ and $G_{\pi }^{(1)}$ make up a pair of
second-class constraints, and the rest of the constraints are first class.
Of course, the second-class sector of the $\mathcal{G}_{ab}^{(1)}$'s is
ambiguous, because the addition of a linear combination of first-class
constraints to $\mathcal{G}_{01}^{(1)}$ will not change the result of the
previous Poisson brackets.

\bigskip \bigskip

We will take advantage of the simplicity of the case-(i) solution %
\eqref{mccosmo} to make some considerations about the relationship between
the extra d.o.f. and the remnant gauge invariance. No local Lorentz
transformation of the tetrad can modify the metric \eqref{metric1}--\eqref{metric2}. 
But it could affect the $f(T)$ dynamics, since it will
produce one of the following results:

\bigskip

\noindent I) It affects the value of $T$; $T$ is no longer a constant, so
the transformed tetrad is not a case-(i) solution [it could be a case-(iii)
solution or not a solution at all].

\bigskip

\noindent II) It affects the value of $T$; $T$ turns to be a different
constant, so the transformed tetrad is another case-(i) solution because the
extra d.o.f. has changed its value.

\bigskip

\noindent III) The (constant) value of $T$ is not affected; the local
Lorentz transformation is a remnant gauge symmetry.

\bigskip \bigskip

To exemplify these situations, let us show two local Lorentz transformations of the tetrad %
\eqref{mccosmo} that do not change the value of $T$ (remnant symmetries):

\bigskip

\noindent 1) A rotation in the $(\mathbf{E}^{2},\mathbf{E}^{3})$ subspace
[the local parameter $\alpha (\mathbf{x})$ is completely free],
\begin{eqnarray}
\mathbf{E}^{0} &=&\cosh \lambda ~\mathbf{dt}+\sinh \lambda ~a(t)~\mathbf{dr}%
~,\ \ \ \ \mathbf{E}^{1}=\sinh \lambda ~\mathbf{dt}+\cosh \lambda ~a(t)~%
\mathbf{dr},  \notag \\
\mathbf{E}^{2} &=&a(t)~r~(\cos \alpha (\mathbf{x})~\mathbf{d\theta }+\sin
\alpha (\mathbf{x})\sin \theta ~\mathbf{d\varphi })~,\ \ \ \ \mathbf{E}%
^{3}=a(t)~r~(-\sin \alpha (\mathbf{x})~\mathbf{d\theta }+\cos \alpha (%
\mathbf{x})\sin \theta ~\mathbf{d\varphi }).  \label{mccosmo1}
\end{eqnarray}

\bigskip

\noindent 2) A boost along the $\varphi $ direction [the local parameter $%
\beta (\mathbf{x})$ cannot depend on $\varphi $ in order to keep the value $%
T=T_{o}$]:
\begin{eqnarray}
\mathbf{E}^{0} &=&\cosh \beta (\mathbf{x})~(\cosh \lambda ~\mathbf{dt}+\sinh
\lambda ~a(t)~\mathbf{dr)+}\sinh \beta (\mathbf{x})~a(t)~r~\sin \theta ~%
\mathbf{d\varphi }~,\ \ \ \ \mathbf{E}^{1}=\sinh \lambda ~\mathbf{dt}+\cosh
\lambda ~a(t)~\mathbf{dr},  \notag \\
\mathbf{E}^{2} &=&a(t)~r~\mathbf{d\theta }~,\ \ \ \ \mathbf{E}^{3}=\sinh
\beta (\mathbf{x})~(\cosh \lambda ~\mathbf{dt}+\sinh \lambda ~a(t)~\mathbf{%
dr)}+\cosh \beta (\mathbf{x})~a(t)~r~\sin \theta ~\mathbf{d\varphi }.
\label{mccosmo2}
\end{eqnarray}

\bigskip

The tetrads \eqref{mccosmo}, \eqref{mccosmo1} and \eqref{mccosmo2} represent
different gauges for the same solution of $f(T)$ gravity, since they share
the value $T=T_{o}$ of the extra d.o.f. They can be distinguished by looking
at non-gauge-invariant quantities, like the torsion vector $T^{\mu }$ and
the axial vector $A_{\mu }\doteq E~\epsilon _{\mu \nu \lambda \rho }~T^{\nu
\lambda \rho }$.\footnote{%
However the divergence of $T^{\mu }$ is a gauge-invariant quantity because
it directly relates to $T$ [see Eq.~\eqref{pseudo}].}

\bigskip \bigskip

On the other hand, a boost along the $r$ direction --the transformation
associated with $G_{01}^{(1)}$-- is able to change the value of the constant
$T_{o}$, thus passing to a different case-(i) solution. In fact, a boost along
the $r$ direction will leave the tetrad \eqref{mccosmo2} unchanged, except
for the replacement
\begin{equation}
\lambda (t,r)~\longrightarrow ~\lambda (t,r)+\gamma (\mathbf{x}),
\end{equation}%
where $\gamma (\mathbf{x})$ is the parameter of the boost. Therefore:

\bigskip

\noindent I) If the parameter $\gamma (\mathbf{x})$ is arbitrary, then $T$
will no longer be a constant. So the transformed tetrad will not be a
case-(i) solution.

\bigskip

\noindent II, III) If the parameter $\gamma (\mathbf{x})$ has the form%
\begin{equation}
\gamma (\mathbf{x})=\Psi (r~a(t))-\frac{1}{4}~r~a(t)~t~\Delta T_{o},
\end{equation}%
then the transformed tetrad will be a case-(i) solution with a different
value of the extra d.o.f.: $T=T_{o}+\Delta T_{o}$ . Thus, the function $a(t)$
will evolve with other effective cosmological and Newton constants.

\bigskip

\section{\protect\bigskip Modifying a higher-order mechanical system with
rotational invariance}

\label{sec:rotinv}

\subsection{Rotationally invariant higher-order Lagrangian}

In this section we will study another toy model and its modification, in
order to show a qualitatively different mechanism for the generation of an
extra degree of freedom. The idea is to mimic the Einstein-Hilbert
Lagrangian which is composed of terms that are invariant under local Lorentz
transformations in the tangent space (they depend just on the metric), but
it exhibits a second-order boundary term to guarantee the invariance under
local diffeomorphisms. So let us introduce a second-order Lagrangian
displaying invariance under local rotations:
\begin{equation}
L=2\left( \frac{d}{dt}\sqrt{z\overline{z}}\right) ^{2}-U(z\overline{z})+A~(%
\ddot{z}~\overline{z}+2~\dot{z}~\dot{\overline{z}}+z~\ddot{\overline{z}}%
)=2\left( \frac{d}{dt}\sqrt{z\overline{z}}\right) ^{2}-U(z\overline{z})+%
\frac{d^{2}}{dt^{2}}\left( A~z\overline{z}\right).  \label{higher}
\end{equation}
The last term is a total derivative which does not enter the Lagrange
equations, so the dynamics is still governed by the equations %
\eqref{motioneq}. However the presence of second derivatives in the
Lagrangian implies the use of Ostrogradsky's procedure to introduce the
Hamilton equations; namely, we have to define momenta associated with both
\textit{canonical variables} $z$ and $Z\equiv \dot{z}$:
\begin{equation}
P_{Z}\equiv \frac{\partial L}{\partial \dot{Z}}=\frac{\partial L}{\partial
\ddot{z}}=A~\overline{z},~~~~~~~~\ ~~~p_{z}\equiv \frac{\partial L}{%
\partial \dot{z}}-\frac{d}{dt}\frac{\partial L}{\partial \ddot{z}}=\frac{1}{z%
}~\frac{d}{dt}(z\overline{z})+A~\dot{\overline{z}},
\end{equation}%
\begin{equation}
P_{\overline{Z}}\equiv \frac{\partial L}{\partial \dot{\overline{Z}}}=\frac{%
\partial L}{\partial \ddot{\overline{z}}}=A~z,~~~~~~~~\ ~~~p_{\overline{z}%
}\equiv \frac{\partial L}{\partial \dot{\overline{z}}}-\frac{d}{dt}\frac{%
\partial L}{\partial \ddot{\overline{z}}}=\frac{1}{\overline{z}}~\frac{d}{dt}%
(z\overline{z})+A~\dot{z}.
\end{equation}%
Thus we get three primary constraints:%
\begin{equation}
G^{(1)}\equiv z~(p_{z}-A~\overline{Z})-\overline{z}~(p_{\overline{z}%
}-A~Z),~~~~~~~~\ ~~G_{Z}^{(1)}\equiv P_{Z}-A~\overline{z},~~~~~~~~\ ~~G_{%
\overline{Z}}^{(1)}\equiv P_{\overline{Z}}-A~z,
\end{equation}
that commute, since
\begin{equation}
\{G^{(1)},G_{Z}^{(1)}\}=0,~~~~~~\{G^{(1)},G_{\overline{Z}%
}^{(1)}\}=0,~~~~~~\{G_{Z}^{(1)},G_{\overline{Z}}^{(1)}\}=0.
\end{equation}%
The canonical Hamiltonian is%
\begin{eqnarray}
H(z,\overline{z},Z,\overline{Z},p_{z},p_{\overline{z}},P_{Z},P_{\overline{Z}%
}) &=&\dot{Z}~P_{Z}+\dot{\overline{Z}}~P_{\overline{Z}}+\dot{z}~p_{z}+\dot{%
\overline{z}}~p_{\overline{z}}-L  \notag \\
&=&\frac{1}{8~z\overline{z}}~\left[ z~(p_{z}-A~\overline{Z})+\overline{z}%
~(p_{\overline{z}}-A~Z)\right] ^{2}+\phi ~U(z\overline{z}).
\end{eqnarray}%
The primary Hamiltonian is%
\begin{equation}
H_{p}=H+u~G^{(1)}+u_{z}~G_{Z}^{(1)}+u_{\overline{Z}}~G_{\overline{Z}}^{(1)}.
\end{equation}%
As already expected, there is not a unique way of writing the canonical
Hamiltonian, due to the presence of constraints. For instance, we can also
write $H=Z~p_{z}+\overline{Z}~p_{\overline{z}}-(2~z~\overline{z})^{-1}\left(
\overline{z}~Z+z~\overline{Z} \right) ^{2}~-2~A~Z~\overline{Z}+U(z\overline{z%
})$. However this apparently simpler form of $H$ will lead to secondary
constraints.\footnote{%
The secondary constraints will be%
\begin{eqnarray*}
G^{(2)} &\equiv &\dot{G}^{(1)}=\frac{Z}{z}~\left( z~p_{z}-\overline{z}%
~Z\right) -\frac{\overline{Z}}{\overline{z}}~\left( \overline{z}~p_{%
\overline{z}}-z~\overline{Z}\right), \\
G_{Z}^{(2)} &\equiv &\dot{G}_{Z}^{(1)}=-p_{z}+A~\overline{Z}+\frac{z~%
\overline{Z}+\overline{z}~Z}{z}, \\
G_{\overline{Z}}^{(2)} &\equiv &\dot{G}_{\overline{Z}}^{(1)}=-p_{\overline{z}%
}+A~Z+\frac{z~\overline{Z}+\overline{z}~Z}{\overline{z}}.
\end{eqnarray*}%
The constraints $G_{Z}^{(1)}$, $G_{\overline{Z}}^{(1)}$, $G_{Z}^{(2)}$, and $%
G_{\overline{Z}}^{(2)}$ are nothing but the definitions of $P_{Z}$, $P_{%
\overline{Z}}$, $p_{z}$, and $p_{\overline{z}}$. Besides $G_{Z}^{(2)}$, $G_{%
\overline{Z}}^{(2)}$ are not linearly independent of \ $G^{(1)}$; in
fact, it is
$G^{(1)}=\overline{z}~G_{\overline{Z}}^{(2)}-z~G_{Z}^{(2)}$. This
nonindependence is an additional ingredient for the right counting
of the degrees of freedom. The secondary constraints should prove to
consistently evolve, which could lead to tertiary constraints.}

\bigskip

The constraints remain zero when the system evolves,
\begin{equation}
\dot{G}^{(1)}=\{G^{(1)},\mathcal{H}_{p}\}=0,~~~~~~\ ~~~~\dot{G}%
_{Z}^{(1)}=\{G_{Z}^{(1)},H_{p}\}=0,~~~~~~\ ~~~~\dot{G}_{\overline{Z}%
}^{(1)}=\{G_{\overline{Z}}^{(1)},H_{p}\}=0.
\end{equation}%
So we have three first-class constraints in a phase space of dimension eight.
The reduced phase space has dimension two, which means one degree of freedom.

\subsection{Modified rotationally invariant higher-order Lagrangian}

Let us deform the theory by replacing the invariant higher-order Lagrangian
with a function of itself,
\begin{equation}
\mathcal{L}=f(L),
\end{equation}%
which is dynamically equivalent to the Jordan-frame representation that
includes an additional dynamical variable $\phi $:
\begin{equation}
\mathcal{L}=\phi ~L-V(\phi ).
\end{equation}%
Again we apply Ostrogradsky's procedure. We will introduce not only the
variable $Z\equiv \dot{z}$, but $\Phi \equiv \dot{\phi}$ as well. Thus the
canonical momenta are
\begin{equation}
\Pi \equiv \frac{\partial \mathcal{L}}{\partial \dot{\Phi}}=\frac{\partial
\mathcal{L}}{\partial \ddot{\phi}}=0,~~~~~~~~\ ~~~\pi \equiv \frac{\partial
\mathcal{L}}{\partial \dot{\phi}}-\frac{d}{dt}\frac{\partial \mathcal{L}}{%
\partial \ddot{\phi}}=0,
\end{equation}%
\begin{equation}
P_{Z}\equiv \frac{\partial \mathcal{L}}{\partial \dot{Z}}=\frac{\partial
\mathcal{L}}{\partial \ddot{z}}=\phi ~A~\overline{z},~~~~~~~~\
~~~p_{z}\equiv \frac{\partial \mathcal{L}}{\partial \dot{z}}-\frac{d}{dt}%
\frac{\partial \mathcal{L}}{\partial \ddot{z}}=\frac{\phi }{z}~\frac{d}{dt}(z%
\overline{z})+\phi ~A~\dot{\overline{z}}-\dot{\phi}~A~\overline{z},
\end{equation}%
\begin{equation}
P_{\overline{Z}}\equiv \frac{\partial \mathcal{L}}{\partial \dot{\overline{Z}%
}}=\frac{\partial \mathcal{L}}{\partial \ddot{\overline{z}}}=\phi
~A~z,~~~~~~~~\ ~~~p_{\overline{z}}\equiv \frac{\partial \mathcal{L}}{%
\partial \dot{\overline{z}}}-\frac{d}{dt}\frac{\partial \mathcal{L}}{%
\partial \ddot{\overline{z}}}=\frac{\phi }{\overline{z}}~\frac{d}{dt}(z%
\overline{z})+\phi ~A~\dot{z}-\dot{\phi}~A~z.
\end{equation}%
Thus, we obtain five primary constraints:%
\begin{equation}
G_{\Pi }^{(1)}\equiv \Pi,~~~~~~~~\ ~~G_{\pi }^{(1)}\equiv \pi,
\end{equation}%
\begin{equation}
G^{(1)}\equiv z~(p_{z}-\phi ~A~\overline{Z})-\overline{z}~(p_{\overline{z}%
}-\phi ~A~Z),
\end{equation}%
\begin{equation}
G_{Z}^{(1)}\equiv P_{Z}-\phi ~A~\overline{z},~~~~~~~~\ ~~G_{\overline{Z}%
}^{(1)}\equiv P_{\overline{Z}}-\phi ~A~z.
\end{equation}%
The Poisson brackets are%
\begin{equation}
\{G^{(1)},G_{Z}^{(1)}\}=0,~~~~~~\{G^{(1)},G_{\overline{Z}%
}^{(1)}\}=0,~~~~~~\{G^{(1)},G_{\Pi }^{(1)}\}=0,~~~~~~\{G^{(1)},G_{\pi
}^{(1)}\}=A~(\overline{z}~Z-z~\overline{Z}),  \label{alg1}
\end{equation}%
\begin{equation}
~\ \ \ \ \ \ \ ~\{G_{Z}^{(1)},G_{\overline{Z}}^{(1)}\}=0,~~~~~~%
\{G_{Z}^{(1)},G_{\Pi }^{(1)}\}=0,~~~~~~\{G_{Z}^{(1)},G_{\pi }^{(1)}\}=-A~%
\overline{z},\ ~~\ ~~  \label{alg2}
\end{equation}%
\begin{equation}
~\{G_{\overline{Z}}^{(1)},G_{\Pi }^{(1)}\}=0,\ ~~~~~~~\{G_{\overline{Z}%
}^{(1)},G_{\pi }^{(1)}\}=-A~z,~~~~~~\{G_{\pi }^{(1)},G_{\Pi }^{(1)}\}=0.
\label{alg3}
\end{equation}%
The canonical Hamiltonian is \footnote{%
Notice that the definition $\Phi \equiv \dot{\phi}$ is necessary to write $%
\mathcal{H}$ in terms of canonical variables.}%
\begin{eqnarray}
&&\mathcal{H}(z,\overline{z},Z,\overline{Z},\phi ,\Phi ,p_{z},p_{\overline{z}%
},P_{Z},P_{\overline{Z}},\pi ,\Pi )=\dot{\Phi}~\Pi +\dot{\phi}~\pi +\dot{Z}%
~P_{Z}+\dot{\overline{Z}}~P_{\overline{Z}}+\dot{z}~p_{z}+\dot{\overline{z}}%
~p_{\overline{z}}-\mathcal{L}  \notag \\
&=&\frac{1}{8~\phi ~z\overline{z}}~\left[ z~(p_{z}-\phi ~A~\overline{Z})+%
\overline{z}~(p_{\overline{z}}-\phi ~A~Z)+2~\Phi ~A~z~\overline{z}\right]
^{2}-\Phi ~A~(\overline{z}~Z+z~\overline{Z})+\phi ~U(z\overline{z})+V(\phi
).
\end{eqnarray}%
However $\Phi \equiv \dot{\phi}$ can be solved from the definitions of $p_{z}
$, $p_{\overline{z}}$ as%
\begin{equation}
\Phi \equiv \dot{\phi}=\frac{-z\ (p_{z}-A~\phi ~\overline{Z})-\overline{z}\
(p_{\overline{z}}-A~\phi ~Z)+2~\phi ~(\overline{z}~Z+z~\overline{Z})}{2~A~~z%
\overline{z}}.  \label{Phi}
\end{equation}%
So, on the primary constraint surface the canonical Hamiltonian can also be 
written as%
\begin{equation*}
\mathcal{H}=-\frac{\phi }{2~z\overline{z}}~(\overline{z}~Z+z~\overline{Z}%
)^{2}+\frac{z\ (p_{z}-A~\phi ~\overline{Z})+\overline{z}\ (p_{\overline{z}%
}-A~\phi ~Z)}{2~z\overline{z}}~(\overline{z}~Z+z~\overline{Z})+\phi ~U(z%
\overline{z})+V(\phi ).
\end{equation*}%
The primary Hamiltonian is%
\begin{equation}
\mathcal{H}_{p}=\mathcal{H}+u~G^{(1)}+u_{Z}~G_{Z}^{(1)}+u_{\overline{Z}}~G_{%
\overline{Z}}^{(1)}+u_{\pi }~G_{\pi }^{(1)}+u_{\Pi }~G_{\Pi }^{(1)}.
\end{equation}%
The consistency equations are%
\begin{equation}
\dot{G}^{(1)}=\{G^{(1)},\mathcal{H}_{p}\}=A~(z~\overline{Z}-\overline{z}%
~Z)~\left( \Phi -u_{\pi }\right),
\end{equation}%
\begin{equation}
~\dot{G}_{Z}^{(1)}=\{G_{Z}^{(1)},H_{p}\}=A~\overline{z}~\left( \Phi -u_{\pi
}\right),
\end{equation}%
\begin{equation}
\dot{G}_{\overline{Z}}^{(1)}=\{G_{\overline{Z}}^{(1)},H_{p}\}=A~z~\left(
\Phi -u_{\pi }\right),
\end{equation}%
where $\Phi $ is given by Eq.~\eqref{Phi}, and
\begin{equation}
\dot{G}_{\pi }^{(1)}=\{G_{\pi }^{(1)},\mathcal{H}_{p}\}=\frac{1+A}{2~z%
\overline{z}}~\left( z\ \overline{Z}+\overline{z}\ Z\right) ^{2}-U(z%
\overline{z})-V^{\prime }(\phi )~-u~A~(\overline{z}~Z-z~\overline{Z})+A~(%
\overline{z}~u_{Z}+z~u_{\overline{Z}}),  \label{consistency}
\end{equation}%
\begin{equation}
\dot{G}_{\Pi }^{(1)}=\{G_{\Pi }^{(1)},\mathcal{H}_{p}\}=0.
\end{equation}%
Thus, no secondary constraints will appear, since the consistency can be
managed by properly choosing the Lagrange multipliers. From the algebra %
\eqref{alg1}, \eqref{alg2} and \eqref{alg3} we recognize one first-class
constraint $G_{\Pi }^{(1)}$ (so $u_{\Pi }$ will be left as a free function
of $t$). Among the other four Lagrange multipliers only two of them seem to
have been determined: $u_{\pi }(t)=\Phi (t)$ (however the evolution of $\Phi
$ is not determined by the Hamilton equations!), and some combination of $u$,$%
~u_{Z}$, and$~u_{\overline{Z}}$ that makes the result \eqref{consistency} zero.
Therefore, four Lagrange multipliers would be left free, which would imply
that the evolution of four of the six variables $z$, $\overline{z}$, $Z$, $%
\overline{Z}$, $\phi $, $\Phi $ are not determined by the Hamilton equations.
The fact that some of the Lagrange multipliers $u$,$~u_{Z}$, $u_{\overline{Z}%
}$, $u_{\pi }$ are not determined by the consistency equations means that
the set $G^{(1)}$, $G_{Z}^{(1)}$, $G_{\overline{Z}}^{(1)}$, $G_{\pi }^{(1)}$
involves first-class constraints. In fact, the matrix%
\begin{equation}
\{G_{i}^{(1)},G_{j}^{(1)}\}=\left(
\begin{array}{cccc}
0 & 0 & 0 & -A(z~\overline{Z}-\overline{z}~Z) \\
0 & 0 & 0 & -A~\overline{z} \\
0 & 0 & 0 & -A~z \\
A(z~\overline{Z}-\overline{z}~Z) & A~\overline{z} & A~z & 0%
\end{array}%
\right)
\end{equation}%
has rank $2$; thus, by combining rows, we can make two of them zero.
Concretely, the constraints can be combined to yield%
\begin{eqnarray}
\mathcal{G}_{1}^{(1)} &=&(z+\overline{z})~G^{(1)}-(z~\overline{Z}-\overline{z%
}~Z)~(G_{Z}^{(1)}+G_{\overline{Z}}^{(1)}),  \notag \\
\mathcal{G}_{2}^{(1)} &=&~z~G_{Z}^{(1)}-\overline{z}~G_{\overline{Z}}^{(1)},
\notag \\
\mathcal{G}_{3}^{(1)} &=&~z~G_{Z}^{(1)}+\overline{z}~G_{\overline{Z}}^{(1)},
\notag \\
\mathcal{G}_{4}^{(1)} &=&G_{\pi }^{(1)}.
\end{eqnarray}%
Thus, the algebra \eqref{alg1}, \eqref{alg2} and \eqref{alg3} is replaced by%
\begin{equation}
\{\mathcal{G}_{i}^{(1)},\mathcal{G}_{j}^{(1)}\}=\left(
\begin{array}{cccc}
0 & 0 & 0 & 0 \\
0 & 0 & 0 & 0 \\
0 & 0 & 0 & -2~A~z~\overline{z} \\
0 & 0 & 2~A~z~\overline{z} & 0%
\end{array}%
\right).
\end{equation}%
Therefore the first-class constraints are $\mathcal{G}_{1}^{(1)}$, $\mathcal{%
G}_{2}^{(1)}$, $G_{\Pi }^{(1)}$, and the second-class constraints are $%
\mathcal{G}_{3}^{(1)}$, $\mathcal{G}_{4}^{(1)}=G_{\pi }^{(1)}$. Then 3+1
degrees of freedom are removed from the canonical variables $z$, $\overline{z%
}$, $Z$, $\overline{Z}$, $\phi $, $\Phi $. Two genuine degrees of freedom
are left. This toy model can be regarded as an analogue of $f(R)$ gravity.
The extra d.o.f. is then analogous to the well-known propagating extra
d.o.f. that results from the trace of the modified Einstein equations in $%
f(R)$ gravity.

\section{Conclusions}

\label{sec:concl} In order to better understand the nature of the extra
degree of freedom in $f(T)$ gravity, we have developed in Sec. \ref%
{sec:toymodel} a toy model endowed with local rotational pseudoinvariance,
that mimics the pseudoinvariance of the TEGR Lagrangian under local Lorentz
transformations of the tetrad field. The nonlinear modification of this
system can then be taken as an analogue of $f(T)$ gravity. We have shown
that the nonlinear modification of a pseudoinvariant system leads to two
different scenarios. In general, one extra d.o.f. should be expected due to
the loss of the local rotational pseudoinvariance in the modified system.
In the so-called case-(i) solutions, the extra d.o.f manifests itself as a
constant of motion affecting the phase of the dynamical variable $z$; it
does not influence the gauge-invariant variable $|z|$, which evolves under
the dictates of the (unmodified) original Lagrangian. The other scenario
relates to the case-(iii) solutions, which make working the modified
dynamics like if they came from an \textit{invariant} Lagrangian [i.e., as 
if they were case-(ii) solutions]. These solutions do not exhibit an extra
d.o.f., but they do exhibit a heavily modified dynamics for the gauge-invariant variables.

The counting of the number of d.o.f., both for the toy model (Sec. \ref%
{sec:toymodel}) and $f(T)$ gravity (Sec. \ref{sec:ftdof}), relies on the
Dirac-Bergmann formalism for constrained Hamiltonian systems, which has been
designed to identify the constraints that generate gauge transformations,
and to separate the spurious d.o.f. We have summarized the qualitative
features of the toy model and TEGR in Table \ref{TM}; the same comparison
between the modified toy model and $f(T)$ gravity is found in Table \ref%
{tmft}. In both models, the distinctive feature is the deformation of the
constraint algebra due to the loss of the pseudoinvariance. As a
consequence, a subset of the Poisson brackets of the constraint algebra
becomes different from zero; however, they could remain zero on some
trajectories of the phase space, which is the key for the branching of the
solutions into case (i) and case (iii). In the case-(i) solutions of $f(T)$
gravity, the scalar torsion $T$ is a genuine d.o.f. that behaves as a
constant of motion. The dynamics of the original gauge-invariant d.o.f. 
(the components of the metric tensor) is dictated by the equations of TEGR
(however the cosmological and Newton constants are shifted as a consequence
of the role of $E$ in the Lagrangian density). In the case-(iii) solutions
the constraint algebra becomes (on-shell) trivial; the extra d.o.f. does not
manifest itself but the metric gets a modified dynamics. Some remnant gauge
symmetry can be left in both cases, since TEGR comes
not with one but six local Lorentz pseudoinvariances (in $n=4$
dimensions). We have exemplified the two different scenarios in the context
of a FLRW flat cosmology. The present analysis strongly suggests the study of
the branching of solutions to $f(T)$ gravity in cases other than the
cosmological one. Some other examples of solutions with $T=$ const in
modified teleparallel gravity have been documented in Refs. \cite%
{Ferraro:2011ks,Boehmer:2011gw,Bohmer:2011si,Nashed:2013mga,Bejarano:2014bca,Nashed:2015lua}. Naturally, the toy model cannot cover all the features of $f(T)$
gravity. The toy model is a mechanical pseudoinvariant system, whereas
$f(T)$ gravity is a field theory.  Because of this reason, there could
 still be room for $f(T)$ solutions exhibiting the extra d.o.f. but
having $T$ different from a constant. The point is that $T$ should not
evolve, as shown by the toy model, so we could consider solutions whose
$T$ only depends on the spatial coordinates. Such solutions could
exhibit both the extra d.o.f. and an effect of modified gravity at the
level of the metric tensor. In this regard, the study of exact wave
solutions to $f(T)$ equations might be a fertile arena for future
research.

Finally, in Sec. \ref{sec:rotinv} we have contributed to deepening the
comparison between $f(R)$ and $f(T)$ gravity by introducing a toy model that
is intended to mimic $f(R)$ gravity. This model is \textit{invariant} under
local rotations; thus its nonlinear modification does not entail the loss
of a local symmetry. However, the model comes with a second-order boundary
term, which will be encapsulated in the function $f$ of the modified
dynamics. Thus, fourth-order Lagrange equations have to be expected for the
modified dynamics, which will cause an extra d.o.f. This toy model is a
good analogue of $f(R)$ gravity because the Einstein-Hilbert Lagrangian is
made of terms that are separately invariant under local Lorentz
transformations of the tetrad (they depend just on the metric). Besides, it
includes an inoffensive second-order boundary term that is needed to achieve
the invariance under local diffeomorphisms. As is well known, $f(R)$ gravity
possesses a {\it propagating} extra d.o.f., whose dynamics is governed by the
trace of the modified Einstein equations. This fact seems to constitute a
remarkable difference when compared with $f(T)$ gravity.

\acknowledgments{R. F. has been funded by Consejo Nacional de Investigaciones
Cient\'ificas y T\'ecnicas (CONICET) and Universidad de Buenos
Aires. M.J.G. has been funded by FONDECYT-ANID Postdoctoral grant No.
3190531. R.F. is a member of Carrera del Investigador Cient\'ifico (CONICET,
Argentina).}

\end{document}